\documentclass[preprint]{aastex}
\usepackage{graphicx,epsfig}
\newcommand\lrest{${\rm L^{rest}}$}
\newcommand\lv{${\rm L_V^{rest}}$}

\newcommand\lb{${\rm L_B^{rest}}$}

\newcommand\zp{$z_{phot}$}
\newcommand\dzp{$\delta z_{phot}$}
\newcommand\dzm{$\delta z_{MC}$}
\newcommand\etal {et al. }
\newcommand\Js{J$_{\rm s}$}
\newcommand\Jsab{J$_{\rm s,AB}$}
\newcommand\Hab{H$_{\rm AB}$}
\newcommand\Ks{K$_{\rm s}$}
\newcommand\Ksab{K$_{\rm s,AB}$}
\newcommand\Ksvega{K$_{\rm s,vega}$}
\newcommand\Hvega{H$_{\rm vega}$}
\newcommand\Jsvega{J$_{\rm s,vega}$}
\newcommand\Ksabtot{K$_{\rm s,AB}^{\rm tot}$}
\newcommand\dz{$\Delta z$}
\newcommand\ang{${\rm \AA}$}
\newcommand\zs{$z_{spec}$}
\newcommand\rh{${\rm R_H}$}
\newcommand\lstar{${\rm L}^*$}
\newcommand\lstarb{${\rm L}^*_B$}
\newcommand\phistar{$\phi^*$}

\begin{document}
\title{A K-band Selected Photometric Redshift Catalog in the HDF-S:
  Sampling the Rest-Frame V-Band to $z=3$\footnote{Based on
    observations with the NASA/ESA \textit{Hubble Space Telescope},
    obtained at the Space Telescope Science Institute, which is
    operated by the AURA, Inc., under NASA contract NAS5-26555.  Also
    based on observations collected at the European Southern
    Observatories on Paranal, Chile as part of the ESO programme
    164.O-0612}}

\author{Gregory Rudnick\footnote{Max-Planck-Institut f\"ur Astronomie,
    K\"onigstuhl 17, Heidelberg, D-69117, Germany, \texttt{grudnick,
      rix@mpia.de}}, Marijn Franx\footnote{Leiden Observatory, PO BOX
    9513, 2300 RA Leiden, Netherlands, \texttt{franx, vstarken,
      pvdwerf, rottgeri, ivo@strw.leidenuniv.nl}}, Hans-Walter
  Rix$^2$, Alan Moorwood\footnote{European Southern Observatory,
    Karl-Schwarzschild Strasse 2, 85748 Garching, Germany,
    \texttt{amoor@eso.org}}, Konrad Kuijken\footnote{Kapteyn
    Institute, Postbus 800, Groningen 9700 AV, the Netherlands,
    \texttt{kuijken@astro.rug.nl}}, Lottie van
  Starkenburg\footnotemark[3], Paul van der Werf, Huub R\"ottgering,
  Pieter van Dokkum\footnote{California Institute of Technology, MS
    105-24, Pasadena, CA 91125, \texttt{pgd@astro.caltech.edu}},
  \& Ivo Labb\'{e}\footnotemark[3]}

\slugcomment{Accepted to the Astronomical Journal, to appear November
  2001}

\begin{abstract}
  
  We present the first results from the \textbf{F}aint
  \textbf{I}nfra-\textbf{R}ed \textbf{E}xtragalactic \textbf{S}urvey
  (FIRES) of the Hubble Deep Field (HDF) South.  Using a combination
  of deep near infrared (NIR) data obtained with ISAAC at the VLT with
  the WFPC2 Hubble Space Telescope data, we construct a K-band
  selected sample which is $50\%$ and $90\%$ complete for \Ksab$\leq
  23.5$ and \Ksab$\leq 22.0$ respectively where the magnitudes are
  measured over a 2\farcs0 diameter aperture.  For $z \leq 3$, our
  selection by the K-band flux chooses galaxies based on wavelengths
  redder than the rest-frame V-band, and so selects them in a way
  which is less dependent on their current star formation rate (SFR)
  than selection in the rest-frame UV.
  
  We developed a new photometric redshift technique which models the
  observed spectral energy distribution (SED) with a linear
  combination of empirical galaxy templates.  We tested this technique
  using 150 spectroscopic redshifts in the HDF-N from the Cohen \etal
  (2000) sample and find \dz$/(1+z)\approx 0.07$ for $z<6$.  We show
  that we can derive realistic error estimates in \zp~by combining the
  systematic uncertainties derived from the HDF-N with errors in
  \zp~which depend on the observed flux errors.  We estimate
  photometric redshifts for 136 galaxies in the HDF-S from the full
  seven-band, $0.3-2.2\mu$m spectral energy distribution.  In finding
  the correct \zp, our deep NIR data is important for breaking the
  redshift degeneracy between templates of identical observed optical
  colors.
  
  The redshift histogram of galaxies in the HDF-S shows distinct
  structure with a sharp peak at $z\approx 0.5$ and a broad
  enhancement at $z\sim 1-1.4$.  We find that $12\%$ of our galaxies
  with \Ksvega$<21$ lie at $z \geq 2$.  While this is higher than the
  fraction predicted in $\Omega_M=1$ hierarchical models of galaxy
  formation we find that published predictions using pure luminosity
  evolution models produce too many bright galaxies at redshifts
  greater than unity.  Finally, we use our broad wavelength coverage
  to measure the rest-frame $UBV$ luminosities \lrest~for $z\leq3$.
  There is a paucity of galaxies brighter than \lv$\geq 1.4\times
  10^{10}h^{-2}L_{\odot}$ at $z\sim1.5-2$, similar to what Dickinson
  (2001b) found for the HDF-N.  However, \zp~is particularly uncertain
  in this regime and spectroscopic confirmation is required.  We also
  note that at $z>2$ we find very luminous galaxies with \lv$\geq
  5\times 10^{10}~ h^{-2}L_{\odot}$ (for
  $\Omega_\mathrm{M}=0.3,~\Omega_{\Lambda}=0.7,~\mathrm{and~H_o}=100~h~km~s^{-1}Mpc^{-1}$).
  Local B-band luminosity functions predict 0.1 galaxies in the
  redshift range $2\leq z \leq 3.5$ and with \lb$\geq 5\times 10^{10}~
  h^{-2}L_{\odot,B}$ but we find 9.  The discrepancy can be explained
  if \lstarb~increases by a factor of 2.4-3.2 with respect to locally
  determined values.  Random errors in the photometric redshift can
  also play a role, and spectroscopic confirmation of the redshifts of
  these bright galaxies are required.

\end{abstract}

\section{INTRODUCTION}

Observational constraints on galaxy formation have improved
dramatically in recent years as large, ground-based telescopes, HST,
and high-efficiency, wide-field instruments, have allowed astronomers,
for the first time, to identify and observe statistically significant
samples of high-redshift galaxies.  With these data, one can address
several important questions: What is the cosmic star formation history
(SFH; e.g. Lilly \etal 1996; Madau \etal 1996)?  What is the mean
stellar age of galaxies as a function of redshift and color (Papovich,
Dickinson, and Ferguson 2001)?  What is the role of dust in galaxy
spectral energy distributions (SEDs; e.g. Thompson, Weymann, \&
Storri-Lombardi 2001; Adelberger \& Steidel 2000; Mobasher \& Mazzei
2000)?  What are the sizes and luminosities of galaxies as a function
of redshift and color (e.g.  Giallongo \etal 2000; Poli \etal 1999;
Lilly 1998; Schade \etal 1996)?  What are the SFHs of galaxies with
different morphologies (Brinchmann \& Ellis 2000)?

The most efficient method to date for detecting and confirming high
redshift galaxies is the Lyman break (LB) technique originally
developed by Steidel \& Hamilton (1992).  Photometric pre-selection by
this method, followed by spectroscopic confirmation at the KECK I \&
II telescopes, has resulted in the discovery of $\sim 900$ galaxies at
$z\gtrsim 2.5$ (e. g. Steidel \etal 1996; Steidel \etal 1999).
Although a powerful tool, this method has two intrinsic limitations:
it is sensitive only to un-obscured galaxies with a high current
star-formation rate (SFR) and it can only find galaxies beyond
$z\gtrsim 2.3$.

A general method for estimating galaxy redshifts from broad-band
photometry is the photometric redshift technique (see e.g.
Contributions in Weymann \etal 1999) of which the LB technique is a
special case.  Connolly \etal (1995) demonstrated for galaxies with
$z\lesssim 1$ that accurate and reliable photometric redshifts
($\langle |z_{spec} -$\zp$|\rangle\sim 0.05$) could be obtained if a
``training set'' with spectroscopic redshifts was available.  When
large spectroscopically confirmed samples with identical photometry
are not available, template fitting can provide alternative redshift
estimates (e.g.  Ben{\'\i}tez 2000; Csabai \etal 2000;
Fern\'andez-Soto, Lanzetta, \& Yahil 1999; hereafter FLY99; Fontana
\etal 1999; hereafter F99; Pascarelle, Lanzetta, \& Fern\'andez-Soto
1998; Giallongo \etal 1998; Sawicki, Lin, \& Yee 1997; Gwyn \&
Hartwick 1996).  There, the most likely redshift for a galaxy is
obtained by comparing multi-band observed colors to the expected
colors of redshifted templates.  With an appropriate choice of
templates - empirical, synthetic, or both - this method allows an
accurate redshift determination across a wide range in $z$ and
independent of the SFH.  As with the LB technique, however, all
photometric methods rely on features in the SED to pin down the
redshift.  Beyond $z \approx 1$ the Calcium H+K ``4000\ang'' break and
the Balmer break are moved into the near infrared (NIR), while the
Lyman break still falls blueward of the atmospheric UV cutoff until
$z\approx 2.3$.  To identify galaxies in the important redshift range
$1\lesssim z\lesssim 2.3$, we need to rely on rest-frame optical
breaks and hence require deep NIR imaging.  Such NIR data can also
detect the Lyman break at very high redshifts ($z\gtrsim 10$).

J, H, and K-band fluxes also allow us to select galaxies at $z\leq 3$
based on their rest-frame V-band light.  Such a selection is less
biased toward galaxies with high star formation rates than a
flux-limited selection in the rest-frame UV.  Indeed, at the present
time, NIR selection is the best practical way to select galaxies by
their stellar mass.  As shown by Kauffmann \& Charlot (1998; hereafter
KC98), the redshift distribution for a sample selected by stellar mass
can serve as a powerful constraint on theories of galaxy formation.
KC98 used semi-analytic models coupled with stellar population
synthesis codes to predict the redshift distribution in K-band
selected samples of differing flux limits.  They found that a generic
prediction of hierarchical models is a lack of K-band luminous
galaxies at high redshift.

We initiated the \textbf{F}aint \textbf{I}nfra-\textbf{R}ed
\textbf{E}xtragalactic \textbf{S}urvey (FIRES; Franx \etal 2000) at
the VLT (Labb\'{e} \etal 2001) to access rest-frame optical
wavelengths over a large range in redshift.  This public dataset
combines some of the deepest optical images from HST with very deep
ground-based \Js H\Ks~data from the Infrared Spectrograph And Array
Camera (ISAAC; Moorwood \etal 1998) at the VLT.  Once complete, this
survey will have accumulated $\approx$192 hours of time with the ISAAC
instrument and $\approx$8 hours of FORS1/2 time to obtain imaging of
both the WFPC2 field of the HDF-S and a mosaic of six WFPC2 fields
covering the $z=0.83$ cluster MS1054--03.  In conjunction with the HST
data, this provides 7-band photometry over an area of $\approx 31$
square arcminutes.  Our unique dataset, coupled with an accurate
photometric redshift technique, will allow us to directly trace the
mass assembly of galaxies regardless of their SFH through a flux
limited selection in the K-band.  Using the redshifts and the observed
SEDs, we can then reconstruct the rest-frame SEDs of galaxies over a
large range in intrinsic luminosity and rest-frame color.

In this paper we present initial results from observations the HDF-S
obtained as part of FIRES.  With these data, we derive photometric
redshifts with accompanying uncertainties and determine the rest-frame
U, B, and V-band luminosities for galaxies with $z\leq3$ in a \Ks-band
selected sample.  Our current data are deep enough to probe galaxies
at $z=2$ with rest-frame luminosities \lrest$\geq 10^{10}~L_{\odot}$.
With our data we place new constraints on the redshift distribution in
the HDF-S for $1<z<2.5$.  In \S2, we present the observations and
data.  In \S3 we discuss our new photometric redshift technique
including a discussion of its reliability.  We show and discuss the
redshift distribution of our sample and the \lrest~values of our
galaxies in \S4.  We summarize in \S5.  We adopt a $\Lambda$-cosmology
throughout  the paper with
$\Omega_\mathrm{M}=0.3,~\Omega_{\Lambda}=0.7,~\mathrm{and~H_o}=100 km
s^{-1} Mpc^{-1}$.  If $h$ is omitted, assume $h=1.0$.

\section{OBSERVATIONS AND DATA}
\subsection{Observations}
We present the first data taken on the HDF-S in the fall of
1999.  The total exposure times were 6.7, 5.7, and 7.5 hours in \Js,
H, and \Ks~respectively.  The field was centered at 22h32m55.03s,
60$^{\circ}$33$'$09\farcs8 (J2000).  All these data were taken in
service mode at the Antu telescope on the nights of 1999 October
21-29, 1999 November 19, and 1999 December 18-19, before its primary
mirror was re-coated.  Despite the reduced sensitivity, the data were
of exceptional quality.  Most of the nights had excellent seeing in
all bands and the combined images had a median image quality of
0\farcs55 (\Js-band), 0\farcs50 (H-band), and 0\farcs50 (\Ks-band).
ISAAC has a pixel scale of of 0\farcs147 pix$^{-1}$ and a field of
almost $150\times150''$ which almost perfectly matches the size of the
WFPC2 field.

Our observing strategy followed established procedures for
ground-based NIR work.  We dithered the images randomly in a 20\farcs0
box to allow the construction of sky frames with minimal object
contamination.  This works well for a field such as the HDF-S which
contains no large, bright objects.  Our exposure times were 120$s$,
120$s$, and 60$s$ split into 4, 6, and 6 integrations for \Js, H, and
\Ks~respectively.

For our optical data, we used the version 2 (Casertano \etal 2000),
reduced, calibrated F300W, F450W, F606W, and F814W WFPC2 data from the
HDF-S.

\subsection{Data Reduction}
\label{datared}

We reduced our ground based images with IRAF\footnote{IRAF is
  distributed by the National Optical Astronomical Observatories,
  which are operated by AURA, Inc. under contract to the NSF.} using
the DIMSUM\footnote{DIMSUM is the Deep Infrared Mosaicing Software
  package developed by Peter Eisenhardt, Mark Dickinson, Adam
  Stanford, and John Ward, and is available via ftp to
  \texttt{ftp://iraf.noao.edu/iraf/contrib/dimsumV2/dimsum.tar.Z}}
package within IRAF and ECLIPSE\footnote{ECLIPSE is a software package
  written by Devillard which is available at
  \texttt{http://www.eso.org/projects/aot/eclipse/}}.  We give a brief
summary of our data reduction below.  For further details see the
presentation of our full dataset (Labb\'{e} et al. 2001).
For each individual science exposure in a given Observing Block (OB),
a sky image was constructed from a maximum of 8 temporally adjacent
images and subtracted from the science frame.  Cosmic rays were
identified from the individual sky-subtracted frames and all the
sky-subtracted frames in a given OB were then aligned and combined.
DIMSUM created a mask marking all pixels belonging to objects by
applying a threshold to the combined image.  Sky-subtraction and
cosmic-ray identification were repeated for the individual frames
using the newly created object mask to exclude object pixels.  We
modified DIMSUM to account for the time-dependent bias in the ISAAC
frames by subtracting the median, on a line-by-line basis, excluding
from the median calculation all object pixels in the object mask.  The
sky-subtracted frames were then flatfielded before the final
registration and combination.  The flatfield images were created from
a time sequence of twilight sky images using the ECLIPSE software.
Individual frames for a given OB were registered and added together
using the imcombine task in IRAF.  The NIR images from all OBs for a
given filter were then combined into a total image.  Finally, we
applied the documented geometric distortion correction to the combined
image while simultaneously interpolating the final NIR images to 4
times the WFPC pixel scale (0\farcs159 pix$^{-1}$).

A weight map was constructed for each NIR passband to reflect the
exposure time at every pixel and hence the noise.  For the HST data we
used the weight maps publically distributed along with the science
frames.  These weight maps were used in all subsequent detection and
photometry steps.

\subsection{Photometric calibration}
\label{calib}

Magnitude zeropoints were derived from standard star observations
taken as part of the normal VLT calibration routine.  For each
standard star, in each filter, and on each night, we measured the flux
in a circular aperture of radius $\sim 3''$ (20 pixels) and used the
magnitude of that star as given in Persson \etal (1998) to establish
our zeropoint for that star.  We derived a nightly zeropoint by
combining all standard star observations in a given night and filter.
By comparing these derived nightly zeropoints to the median zeropoints
over all nights we identified non-photometric nights.  We used the
mean of the zeropoints on the photometric nights to determine the
zeropoint for each bandpass.  The uncertainties in the final
zeropoints were $\sim0.02$.  Using these zeropoints, we derived the
magnitudes of bright stars in the field for the OB's on the
photometric nights, and used them to calibrate the final combined and
distortion corrected image.  All magnitudes in this paper are given in
the AB system unless stated explicitly otherwise.  For the NIR data,
the adopted transformations from the Vega system to the AB system are
taken from Bessell \& Brett (1988; \Jsvega~= \Jsab~- 0.90, \Hvega~= \Hab~-
1.37, \Ksvega~= \Ksab~- 1.88).

In our final reduced images, the $10\sigma$ magnitude limits in a
2\farcs0 circular aperture are $m_{AB} =$ 23.8, 23.0, and 23.2 in Js,
H, and Ks respectively.  The $3\sigma$ limits are $m_{AB} =$ 25.1,
24.4, and 24.5.  Our data are $\sim$ 0.25, 0.1, and 0.2 magnitudes
deeper in \Js, H, and \Ks~respectively than the data on the HDF-N
taken at the Kitt Peak 4-meter with the IRIM camera in April of 1996
(Dickinson \etal 2001a).  The F110W and F160W HDF-N NICMOS data
(Dickinson \etal 2001b) goes 1.1 and 1.9 magnitudes deeper
respectively than our \Js~and H data.  In the HDF-S our data are
$\sim$ 1, 2.1, and 2.1 magnitudes deeper in \Js, H, and
\Ks~respectively than the EIS data from da Costa et al. (1998).

\subsection{Object Detection and Photometry}
\label{detect}

Our first goal is to construct a \Ks-band flux-limited catalog of
objects.  We used the SExtractor software (Bertin \& Arnouts 1996) to
detect objects from the final \Ks~image, using the \Ks-band weight
image.  Faint objects are detected against a noisy background after
convolving the image with a kernel representing the typical expected
object size.  Because SExtractor allows only one convolution kernel
per detection pass, we must optimize the detection for a particular
object size, biasing ourselves against faint objects of very different
sizes.  We choose a 0\farcs48 FWHM Gaussian convolution kernel,
extending over $0\farcs8\times 0\farcs8$ which represents the size of
the seeing disk.  As in all deep surveys, deblending of overlapping or
close object pairs is difficult and to some extent subjective.  An
ideal deblending algorithm will not ``oversplit'' single galaxies with
knotty internal structure, but will split close groupings of separate
galaxies.  We settled on a single set of deblending parameters that
nearly eliminate the over splitting of galaxies: DEBLEND\_NTHRESH = 32,
DEBLEND\_MINCONT = 0.0002.  These parameters set the number of
deblending sub-thresholds and the minimum contrast needed to deblend
two objects, respectively.

To obtain consistent photometry across the full seven bands, we need
to account for the vastly different pixel scales and resolutions
between our space-based and ground-based images.  To this end, we
first resampled all of the data to the same pixel scale, fitted the
PSF in the NIR images with a double Gaussian, whose equally weighted
components have FWHM$=0\farcs38$ and FWHM$=0\farcs75$ respectively,
and convolved this with the optical data.  To measure colors over
identical angular scales in each band, we choose to measure the fluxes
of all objects in a fixed 2\farcs0 diameter aperture whose position
was chosen from the \Ks-band image.  For the largest objects this
aperture misses some flux, but this choice lessens the chance of
measuring flux from two separate objects.  Still, there are 6 pairs of
objects whose 2\farcs0 apertures overlap (IDs=98,99; 117,127; 187,188;
354,364; 372,373; 397,398).  For some of these objects, the flux
measurements of the galaxy might be strongly affected by the light from
its nearest neighboor.  In calculating the flux errors in all the
images, we used the weight images discussed in (\S\ref{datared}).

We used SExtractor to detect objects using a detection threshold of
0.8 times the standard deviation of the background.  The relative
strength of the background at each pixel was given by the \Ks-band
weight image.  For an object to enter the initial catalog we required
that a minimum of 5 contiguous pixels lie above the detection
threshold.  From the resulting initial catalog of 615 objects detected
in the \Ks-band image, we constructed a catalog optimized for
photometric redshift estimates based on three criteria.  1) To
homogenize the data quality, the value of the exposure time weight
must exceed 0.5 and 0.25 for the VLT and HST images respectively (this
cut reduces our total usable image area to 4.3 arcmin$^2$). 2) To
differentiate between stars and galaxies, we examined the FWHM and
magnitude of objects in the F814W image.  Objects were identified as
stars if they satisfied either of the following two criteria: FWHM
$<6$ pixels and F814W$_{\rm AB} <27$ or FWHM $<15$ pixels and
F814W$_{\rm AB}<22$.  The second of these criteria was used to
eliminate saturated stars. 3) To limit ourselves to magnitudes where
the completeness is greater than $50\%$, we require that the object
must have a total magnitude (the ``AUTO'' magnitude from SExtractor
with a minimum 2\farcs0 diameter) of \Ksabtot$\leq 23.5$, roughly a
$6\sigma$ detection (see \S\ref{compl}).  The exposure time criterion
reduced the initial catalog to 316 objects and the removal of all
point sources in the F814W image left 293.  Of these, 136 objects had
\Ksabtot$\leq 23.5$ and were entered into our final catalog (see Table
1).  The \Ks-band image is shown in Figure~\ref{image} along with all
136 objects and their ID numbers from the final catalog.  All flux
measurements are summarized in Table 1.

\subsection{Completeness}
\label{compl}

\begin{figure}
\scalebox{0.8}{\includegraphics{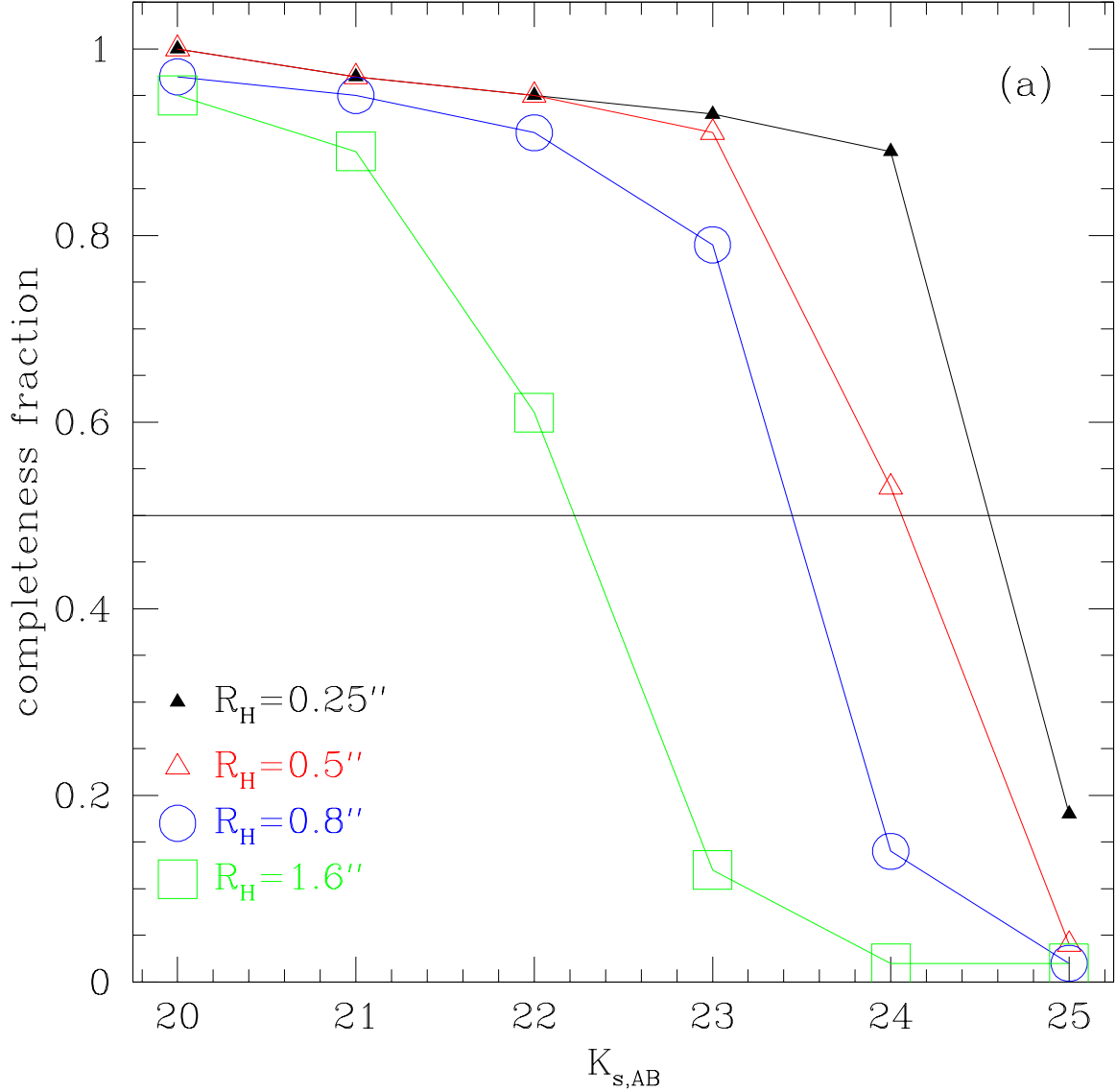}}
\caption{
  Estimates of the \Ks-band completeness limit.  \textbf{a)}
  Completeness against \Ksab~magnitude for galaxies with an
  exponential profile and an axis ratio, $b/a=0.8$.  Different points
  represent different galaxy half-light radii \rh.  Note how the
  completeness dependents greatly on the object size.  \textbf{b)}
  Completeness vs. \Ksab~magnitude, at the typical faint object radius
  of \rh$=$0\farcs8, for three different profile shapes.  The
  completeness is relatively insensitive to the exact profile shape.
  In both plots, the horizontal line shows the $50\%$ completeness
  limit.  }
\label{complete}
\end{figure}

The issue of completeness must be addressed in every survey for faint,
extended objects.  The detectability of an object depends not only on
its apparent magnitude, but also on its morphology and mean surface
brightness.  The detection algorithm used by SExtractor looks for
continuously connected pixels above a certain threshold with respect to
the background.  Relatively bright objects of low surface brightness
may be missed by this technique.  To understand our detection
completeness we added objects to the \Ks-band image and then
determined how successful we were at detecting them.  We constructed
three different types of model objects: An elliptical galaxy with a de
Vaucouleurs profile and an axis ratio of $b/a=0.7$ and two exponential
galaxies with $b/a=0.4$ and $b/a=0.8$.  For each of these three
profile types, we made a magnitude grid of \Ksab=20, 21, 22, 23, 24,
and 25 and a size grid of \rh$=$0\farcs25, 0\farcs5, 0\farcs8, and
1\farcs6 where \rh~is the half-light radius.  For each profile type,
magnitude, and size, we convolved the synthetic galaxy images with the
seeing (see \S2.4) and inserted about 50 such objects into the
\Ks~image at simple grid positions.  We then ran SExtractor on the new
image and counted how many of the model objects were detected for each
set of parameters and how well these parameters (apparent magnitude
and size) were recovered.  Figure~\ref{complete}a shows how the
completeness depends on surface brightness, parameterized by both
input magnitude and size, for a given profile shape and axial ratio.
For a fixed size Figure~\ref{complete}b shows how little completeness
changes with profile shape.

To asses the actual 50\% completeness limit for our sample we must
select size parameters most applicable to galaxies near our flux
limit.  To map the input sizes used in Figure~\ref{complete} to the
sizes returned by SExtractor for the model images, we compared, for
different magnitudes, \rh~to
\begin{equation}
  R_{out} = \sqrt{R_{kron}^2 - R_{seeing}^2}~.
\end{equation}
Here $R_{kron}$ is the Kron radius (Kron 1980) calculated by
SExtractor, and $R_{seeing}$ is the FWHM/2 of the actual observations.
At the faintest level where we could both retrieve the input magnitude
and also see a defined relation between input and output size
(\Ksab$\sim22$), we measured that objects had a typical $R_{kron}$ of
0\farcs6.  Using our input-output size relations, averaged over
profile type, we associated this measured radius with an intrinsic
\rh~of 0\farcs8.  As a choice of profile type we conservatively chose
the curve for which we are least complete, the exponential disk with
$b/a=0.8$ (see Figure~\ref{complete}b).  Using this curve (see
Figure~\ref{complete}a), we established a $50\%$ completeness limit at
\Ksab$= 23.5$ and note that we are $90\%$ complete for \Ksab$< 22.0$.
For this flux limit our conclusions are insensitive to completeness
corrections, and so we make no such corrections.

\section{PHOTOMETRIC REDSHIFTS}
\subsection{Template Choice}

The next step in the analysis is to convert the flux measurements of
objects in the seven bands into an estimate of their redshift.  We
estimate the redshifts of our galaxies by modeling their rest-frame
colors by a combination of empirical spectral templates.  We used
Hubble type templates E, Sbc, Scd, and Im from Coleman, Wu, \& Weedman
(1980; hereafter CWW) and the two starburst templates with a low
derived reddening, designated SB1 and SB2, from Kinney \etal (1996).
For the two starburst templates, the color excess E(B-V) with respect
to the expected colors of an unreddened galaxy is $\leq 0.10$ and
$0.11\leq$ E(B-V) $\leq 0.21$ respectively.  These templates are
needed because many galaxies even in the nearby universe have colors
bluer than the bluest CWW templates and the inclusion of SB1 and SB2
significantly improves the photometric redshift estimate (see also
Sawicki, Lin, \& Yee 1997; Ben{\'\i}tez 2000).

To extend the CWW and starburst templates from their published
short-wavelength limits (1400\ang~and 1232\ang~respectively) to below
the Lyman break, we extrapolated blueward a power law fit to the
1400-1800\ang~and 1240-1740\ang~wavelength ranges, respectively.  To
account for intervening absorption from neutral cosmic hydrogen, we
applied to all our template spectra, the redshift dependent cosmic
mean opacity taken from Madau (1995).  We accounted for the internal
hydrogen absorption of the galaxy by setting the flux blueward of
912\ang~to zero.  To extend the templates to the IR, we used the
stellar population synthesis code of Bruzual \& Charlot (2001).  We
constructed NIR SED extensions for each template by using the stellar
population ages, star formation timescales, and initial mass functions
for each template Hubble type from Pozzetti, Bruzual, \& Zamorani
(1996; see Table 2).  We verified that these SEDs matched the optical
colors of our templates.

In addition to the ``natural'' reddening already included in the
templates, additional reddening may be present.  We will examine the
effect of reddening on the determination of \zp~in Labb\'{e} et al.
(2001).

\subsection{Template Based Estimates of the Redshift}
\label{method}

We cannot assume {\it a priori} that distant galaxies have SEDs
identical to any one of our empirical SEDs.  In fact, even within a
single galaxy there may be spatial variations in the stellar
populations and SFR.  Our goal is to fit the observed flux points as
well as possible with minimal assumptions about the galaxy's SFH.
Therefore, we attempt to model the observed SED by a \textit{linear
  combination} of redshifted templates.  We estimate the likelihood
that a galaxy lies at a given redshift by calculating
\begin{equation}
  \chi^2(z)=\sum_{i=1}^{N_{filter}}\Biggl[\frac{F_i^{data} -
    F_i^{model}}{\sigma_i^{data}}\Biggr]^2~,
\end{equation}
where $F_i^{data}$ is the measured flux value, in units of
$f_{\lambda}$, in the $i$th color bandpass, $\sigma_i^{data}$ is its
associated $1\sigma$ uncertainty and
\begin{equation}
  F_i^{model} = \sum_{j=1}^{N_{template}} C^j\times F_i^j(z),
\end{equation}
where the $F_i^j(z)$ is the flux of the $j$th template, redshifted to
$z$, adjusted for intervening cosmic hydrogen absorption, and
integrated over the transmission curve of the $i$th filter.  For every
redshift we determine the non-negative coefficients $C^j$ which
minimize $\chi^2$ and the most likely photometric redshift, \zp, which
is the minimum of $\chi^2(z)$.  To determine how our photometric
errors propagate to errors in \zp, we performed a Monte-Carlo
simulation where, for each object, we create 200 synthetic photometry
measurements distributed like a Gaussian around the observed flux,
with a width $\sigma = \sigma_i^{data}$.  For each object's
Monte-Carlo set of fluxes, we determined, individually, the values of
\zp~and calculated its 68\% confidence limits \dzm~from the resulting
distribution.  We added a systematic error component in
\S\ref{mismatch} to obtain the final error estimate \dzp.  From this
point on, all values of \zp~will refer to those calculated directly
from the catalog data.  The values of \zp~and \dzp~are given in Table
3.

\subsection{Comparison With Spectroscopic Redshifts}
\subsubsection{The Hubble Deep Field North}

\begin{figure}
\scalebox{0.8}{\includegraphics{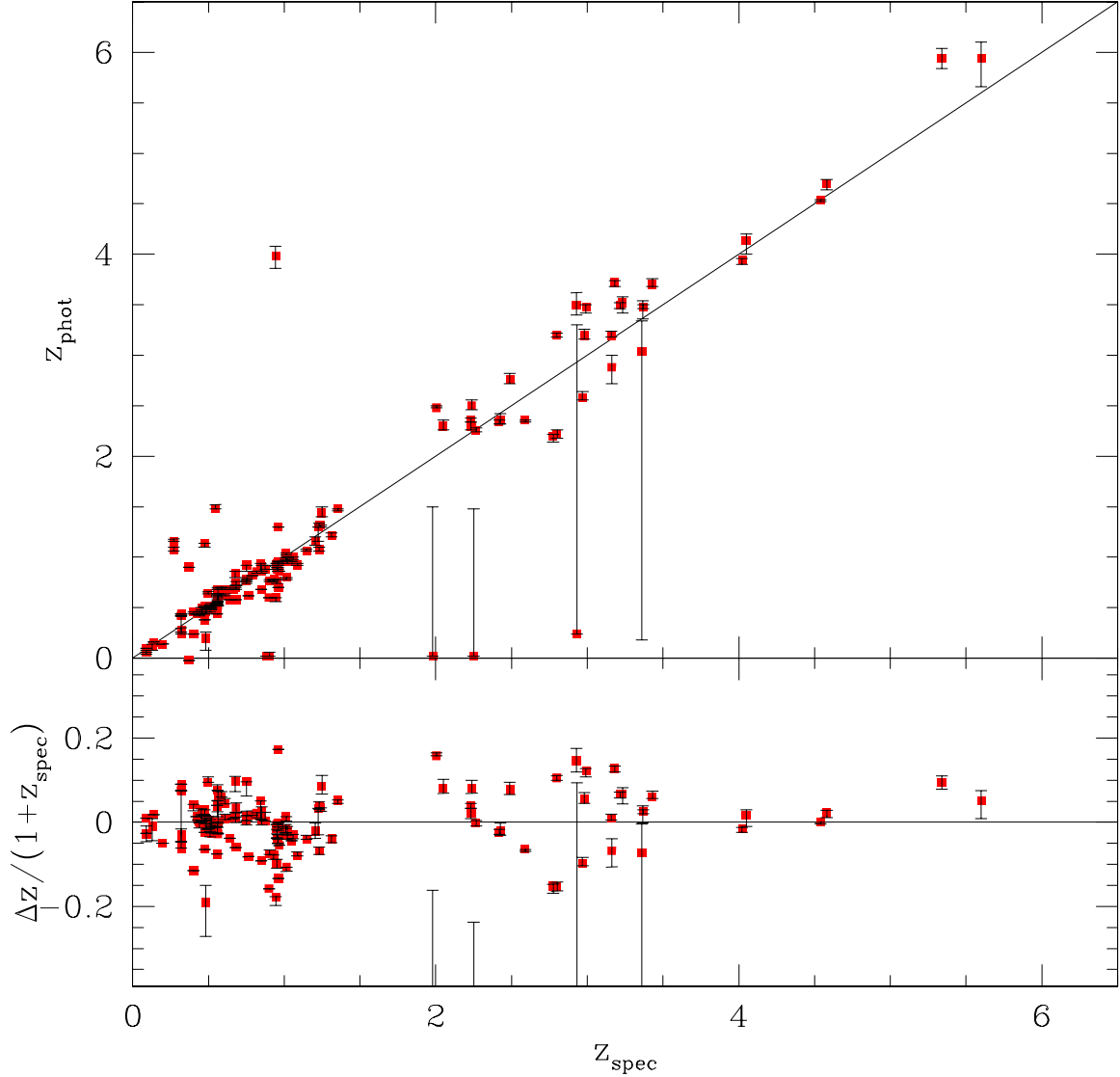}}
\caption{
  A comparison of \zp~to \zs~for objects in the WFPC2 field of the
  HDF-N.  The error bars are derived from our Monte-Carlo simulations.
  The top panel shows a direct comparison between \zp~and \zs.  The
  diagonal line corresponds to a one-to-one relation to guide the eye.
  The bottom panel shows how \zs~relates to the difference between
  \zp~and \zs~normalized by $1+z_{spec}$.  The agreement is excellent
  for \zs$<6.0$ with only $\lesssim 3\%$ of the sample having
  $\left|z_{spec}-z_{phot}\right| > 1.0$ and with \dz$/(1+z)=0.07$.
  The Monte-Carlo errors serve as a good indication of possible
  catastrophic failures of the \zp~determination.}
\label{hdfn}
\end{figure}

We gauged the precision and accuracy of our photometric redshift
technique against spectroscopic redshifts, using the data set provided
by Cohen \etal (2000) on the HDF-N.  This field has optical data from
HST (Williams \etal 1996) and JHK data from the IRIM camera on the
Kitt Peak 4-meter telescope taken by Dickinson \etal (2001a) in April
of 1996.  Using the photometry of FLY99 we derive the photometric
redshifts of all the F814W selected objects in the HDF-N using our
code.  There are a total of 150 objects common between the Cohen \etal
spectroscopic sample and the FLY99 photometric sample.  The comparison
between our photometric redshifts \zp~and the spectroscopic redshifts
\zs~for this sample is shown in Figure~\ref{hdfn}.  The redshift error
bars here are those calculated from the Monte Carlo simulation
\dzm~(see \S\ref{method}).  We choose for our measure of photometric
redshift accuracy
\begin{equation}
  \Delta z = \left|z_{spec}-z_{phot}\right|.
\end{equation}
Our mean value is \dz$\approx0.14$ for $z \leq 1.5$ and
\dz$\approx0.44$ for $z > 1.5$.  We also note that the value
\dz$/(1+z)$ is nearly constant with redshift with
\dz$/(1+z)\approx0.09$ for the whole sample.  This was first noted by
FLY99 and likely stems from the effect that the filter spacing is
roughly constant in $ln(\lambda)$ and the redshift determination is
equivalent to finding a constant shift $ln(1+z)$ for the spectrum if
it is expressed as a function of $ln(\lambda)$.

We note that there are a few objects ($\lesssim 3\%$) for which
\zp~and \zs~are greatly different, in part because there appear to be
galaxies whose SEDs cannot be represented by our template set.  Also,
Fern\'andez-Soto \etal (2001; hereafter FS01) suggested that five of
the published spectroscopic redshifts may be in error.  One of these
objects (FS01 ID number HDF36441\_1410) has a \zs=2.267 and is found
by FS01 to have \zp=0.01.  We however find \zp=2.26, in excellent
agreement with the spectroscopic redshift.  Eliminating HDF36441\_1410
causes almost no change in \dz~or \dz$/(1+z)$ for $z > 1.5$.  Four
objects remain\footnote{Fern\'andez-Soto \etal (2001) ID numbers:
  HDF36396\_1230, HDF36494\_1317, HDF36561\_1330, and HDF36569\_1302}
for which we found that our \zp~values do not agree well with the
published \zs~values.  These objects all lie at \zs$< 1$.  When
eliminating these four objects, we found that \dz~decreased to
$\approx 0.10$ for $z \leq 1.5$.  With these four objects removed the
mean \dz$/(1+z)$ for the redshift range $z<6$ is 0.07.  There are
three objects with \zs=2.931, 2.250, and 1.980 which are not flagged
by FS01 as having an incorrect spectroscopic redshifts (FS01 IDs
HDF36478\_1255, HDF36446\_1227, and HDF36498\_1415) for which we find
\zp=0.024, 0.02, and 0.02 and for which FS01 find \zp=0.26, 2.47, and
1.64.  In all three of these cases, \dzm~is large and so in general,
may provide a good indicator of discrepant \zp~values.

To test the importance of the NIR data in determining the correct
redshift, we compare the accuracy of \zp~in the HDF-N as derived with
and without NIR data.  The NIR data is excluded from the fit by
setting the error term to infinity in the $\chi^2$ sum.  For
\zs$\leq1.5$ the advantage of the NIR data is obvious, with the mean
value of \dz~increasing from 0.10 to 0.21 when the NIR data is not
included.  For two galaxies (FS01 ID HDF36498\_1415, HDF36446\_1227)
with \zs$ = 1.98$ and \zs$ = 2.25$ however, excluding the NIR data
causes \zp~to change from 0.20 to 2.24 and from 0.20 to 2.20
respectively.  The original estimates were obviously wrong.  In both
of these cases, the inclusion of the NIR data forces the code to
incorrectly identify a Lyman break, just entering the F300W band, as a
rest-frame optical break.  When leaving out these two galaxies, \dz~at
\zs$> 1.9$ remains unchanged by the omission of the NIR data.  We
should expect that the NIR data should improve the accuracy of the
redshifts, but it is possible that the flux errors in the NIR have
been underestimated by FLY99 and that these data may overly contribute
to the $\chi^2$.  Unfortunately, the importance of the NIR data cannot
be assessed in the redshift range $1.3 < z < 2$ due to the lack of
spectroscopic redshifts.  In this regime however, only rest-frame
optical breaks are observable and the NIR data is needed to constrain
their position.

\subsubsection{The Hubble Deep Field South}
\begin{figure}
\scalebox{0.8}{\includegraphics{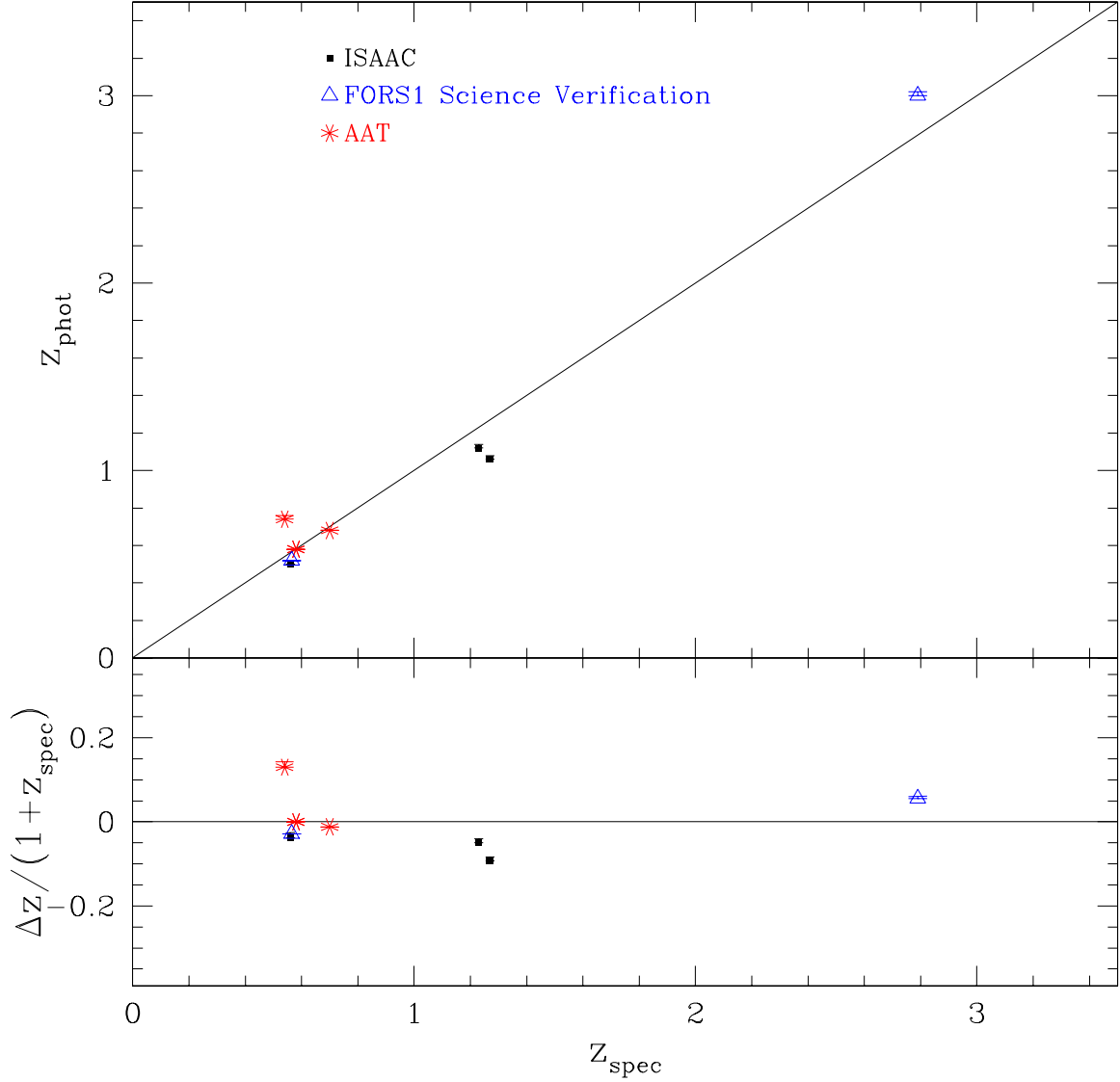}}
\caption{
  A comparison of \zp~to \zs~for objects in the WFPC2 field of the
  HDF-S.  The explanation of this figure is identical to
  Figure~\ref{hdfn}.  \zp~and \dzp~are derived from a Monte-Carlo
  simulation using the formal photometric errors.  Two objects with
  \zs$=0.58$ measurements from the AAT both have values of
  \zp$=0.58$.}
\label{hdfs}
\end{figure}

For the HDF-S we selected all the objects in our catalog with
publically available spectroscopic redshifts.  These include five
objects detected by ISOCAM (Rigopoulou \etal 2000) with spectroscopic
redshifts from ISAAC, two objects from the FORS1 commissioning data
(Cristiani \etal 2000), and four objects with unpublished spectra
taken with the Anglo Australian Telescope (AAT; Glazebrook \etal 2001; hereafter G01; available at
\texttt{http://www.aao.gov.au/hdfs/}), all of which lie in our area
with ``good photometry''.  Two of the objects from G01 also had
spectra from Rigopoulou \etal (2000) which yielded identical values of
\zs.  The comparison of our \zp~to \zs~for these objects is shown in
Figure~\ref{hdfs}.  We find excellent agreement between \zp~and
\zs~with \dz$\approx0.05,0.18$ for $z \leq 1.0$ and $z > 1.0$
respectively.

\subsubsection{Template Mismatch and Redshift Uncertainties}
\label{mismatch}

The photometric redshift error bars derived solely from the Monte
Carlo simulation described in \S\ref{method} significantly
underestimate the true variance of \zp~when compared to \zs.  This is
because the galaxies with spectroscopic redshifts are among the
brightest galaxies in our sample, with very small formal flux errors.
The resulting range of statistically acceptable redshifts and SEDs is
very small and our coarse and finite set of templates significantly
distorts \zs, but is not modeled by our Monte Carlo estimates.  At the
faint end, the photometric errors become large, and dominate the
uncertainty in the redshift, implying realistic error estimates.  Both
effects were noted by FLY99.

We first attempted to compensate for this ``template mismatch'' in the
bright galaxies by using a minimum photometric error of 10\% chosen
such that our Monte Carlo error bars reflect the deviation of \zp~from
\zs.  By introducing a minimum flux error we lessen the relative
contribution of the high S/N HST data points to the $\chi^2$ budget -
which in turn changes the formal best-fit redshift.  A detailed
examination of this effect in the HDF-N and HDF-S data showed that
while this minimum photometric error brought the \zp~values into
statistical agreement with \zs, the actual best-fit values of
\zp~agreed worse with \zs~than when using the formal photometric
errors.  In fact, $\sim20\%$ of the galaxies in both the HDF-N and
HDF-S have \zp~values calculated with the formal flux errors which lie
outside the $68\%$ confidence limits allowed with the boosted flux
errors.

Hence, a proper estimate of the uncertainty in \zp~must take into
account both systematic uncertainties arising from template mismatch
and the uncertainties in \zp~which result from the photometric errors.
We define the total uncertainty in \zp~as
\begin{equation}
  \delta z_{phot} \equiv \sqrt{\langle \left| \Delta z\right| \rangle^2 
    + \delta z_{MC}^2},
\end{equation}
where $\langle \left| \Delta z\right| \rangle$ is the value of $(1
+$\zp$)$ times the mean value of \dz$/(1+z)=0.07$ as derived from the
HDF-N and \dzm~is again the $68\%$ confidence limit of \zp~as derived
from the Monte-Carlo simulation.  Note that \dzm~need not be symmetric
around \zp~and that we add \dz~in quadrature separately for the upper
and lower error bars.  Again, the values of \dzp~are listed in Table
3.

In addition to providing realistic error bars it is also informative
to flag objects with secondary minima in their $\chi^2(z)$
distributions.  Although some secondary minima in $\chi^2(z)$ are
reflected by large values of \dzm, some objects with small \dzm~may
have a finite fraction of the Monte-Carlo realizations which end up at
a rather different redshift.  In fact some of the objects with large
\dz~in the HDF-N have secondary minima close to \zs~which are too
small to be included in \dzm.  In addition to supplying the error bars
which define the range of a galaxy's most likely redshifts, we flag in
Table 3 the 12 objects for which $\geq1\%$ of the Monte-Carlo
realizations lie greater than unity in redshift away from \zp.

\section{Results}

In the section below, we use our estimates of \zp~to examine the
redshift distribution of galaxies in the HDF-S.  We also use our
estimate of \zp, coupled with our broad wavelength coverage, to
determine the rest-frame optical SEDs and luminosities of our galaxies
across a wide range in redshift.

\subsection{SED Fits}

\begin{figure}
\scalebox{0.8}{\includegraphics{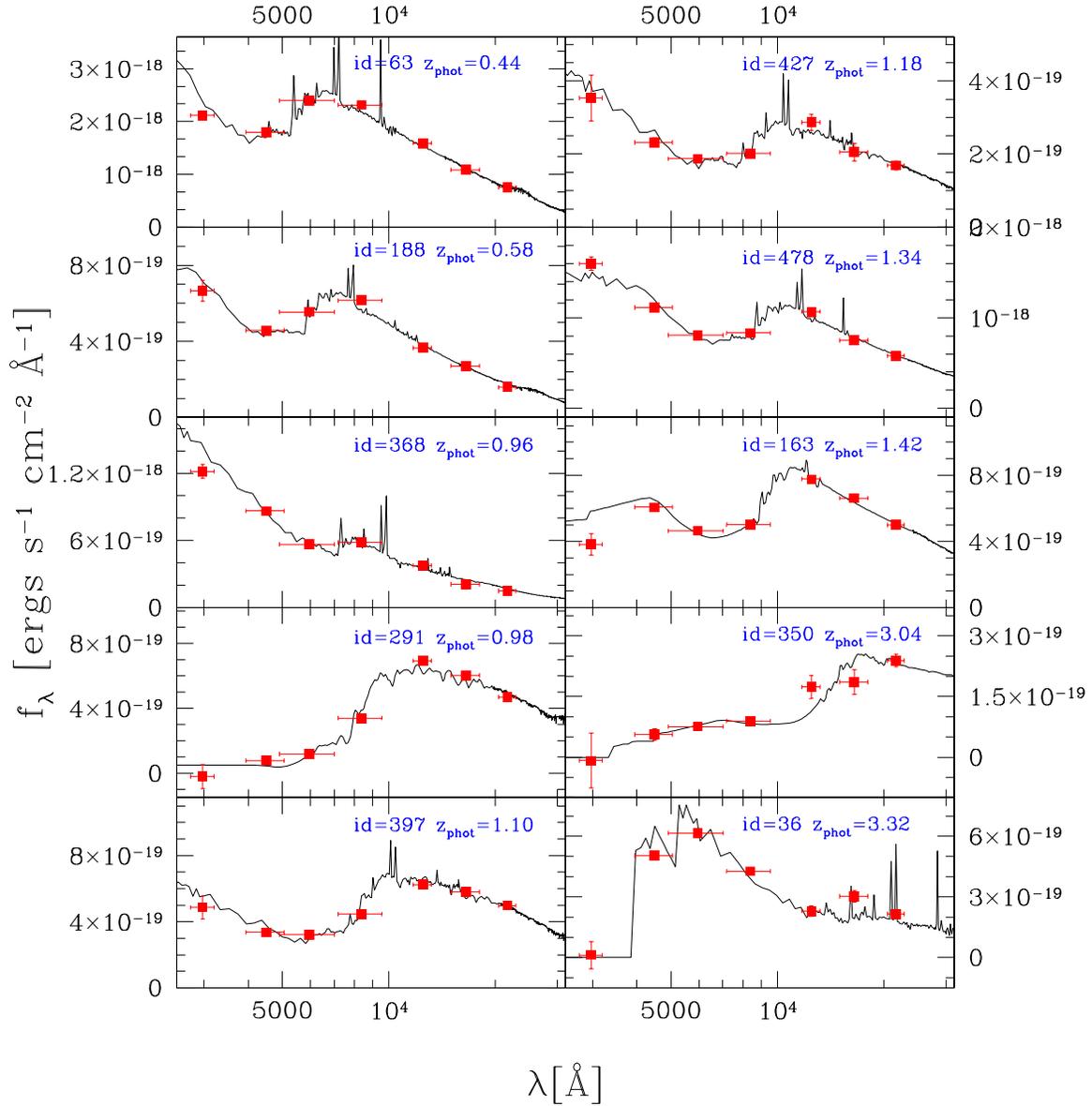}}
\caption{
  A sample of template fits to photometric data for 10 objects in the
  HDF-S.  The measured \zp~increases down and to the right.  In
  addition to blue, star-forming galaxies, there are many galaxies at
  $z>1$ with strong Balmer or 4000\ang~breaks.  }
\label{fit_examples}
\end{figure}

\begin{figure}
\scalebox{0.8}{\includegraphics{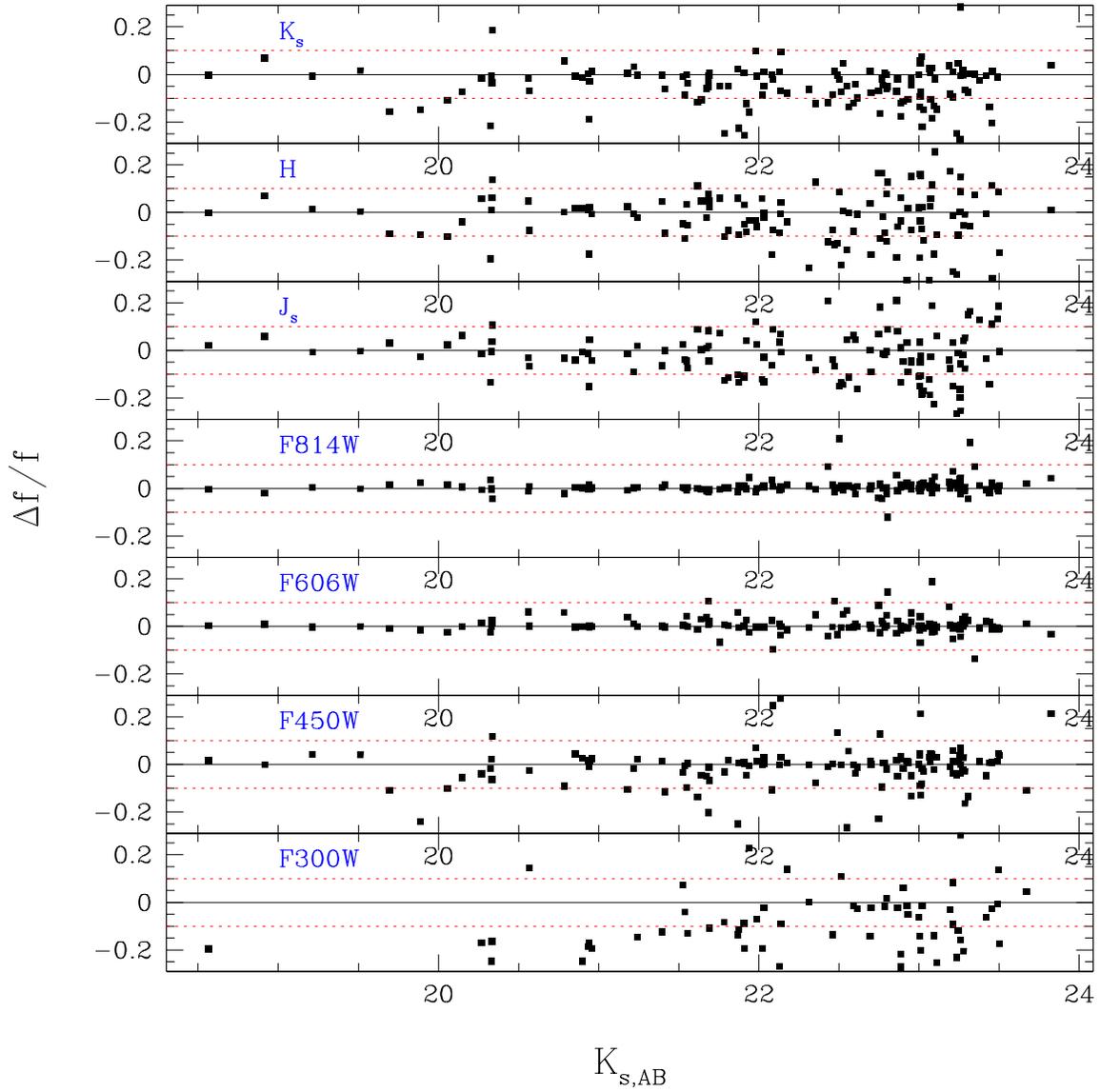}}
\caption{
  The fractional difference between the observed and model fluxes at
  the best-fit redshift as a function of \Ksab.  The horizontal dotted
  lines are at $\pm10\%$ to guide the eye.  The high signal-to-noise
  of the F814W and F606W data forces the best-fit SED to always pass
  close to these points. }
\label{diff_mod}
\end{figure}

In Figure~\ref{fit_examples} we show 10 examples of SED fits to the
seven-band photometry ($0.3-2.2\mu$m) for galaxies in the HDF-S.  In
our analysis of \Ks-band selected galaxies in the HDF-S we find
galaxies with a range of SEDs at all redshifts $0<z<3$ with SED shapes
ranging from very blue starburst templates to earlier Hubble type
templates.  As is shown in Figure~\ref{fit_examples} we also find
galaxies with strong rest-frame 4000\ang~breaks or Balmer breaks at
$z>1$.  These breaks signal that the rest-frame optical light is
dominated by stars at least as old as $A$ stars.  Note that the small
flux errors of the F606W and F814W data force the best-fit SED at any
redshift to pass always through these two points.  This is best shown
in Figure~\ref{diff_mod} where, for each of our 136 galaxies, we plot
the fractional difference between the measured flux and the model flux
of our best-fit SED as a function of \Ksab.  At all magnitudes, the
residuals are lowest in the F606W and F814W bands even if they are
very large in other bands.  This plot is also useful for finding
systematic differences between the SEDs and the data.  For example, it
is seen that the best-fit SED slightly overpredicts the F300W flux at
all magnitudes.

\begin{figure}
  \scalebox{0.8}{\includegraphics{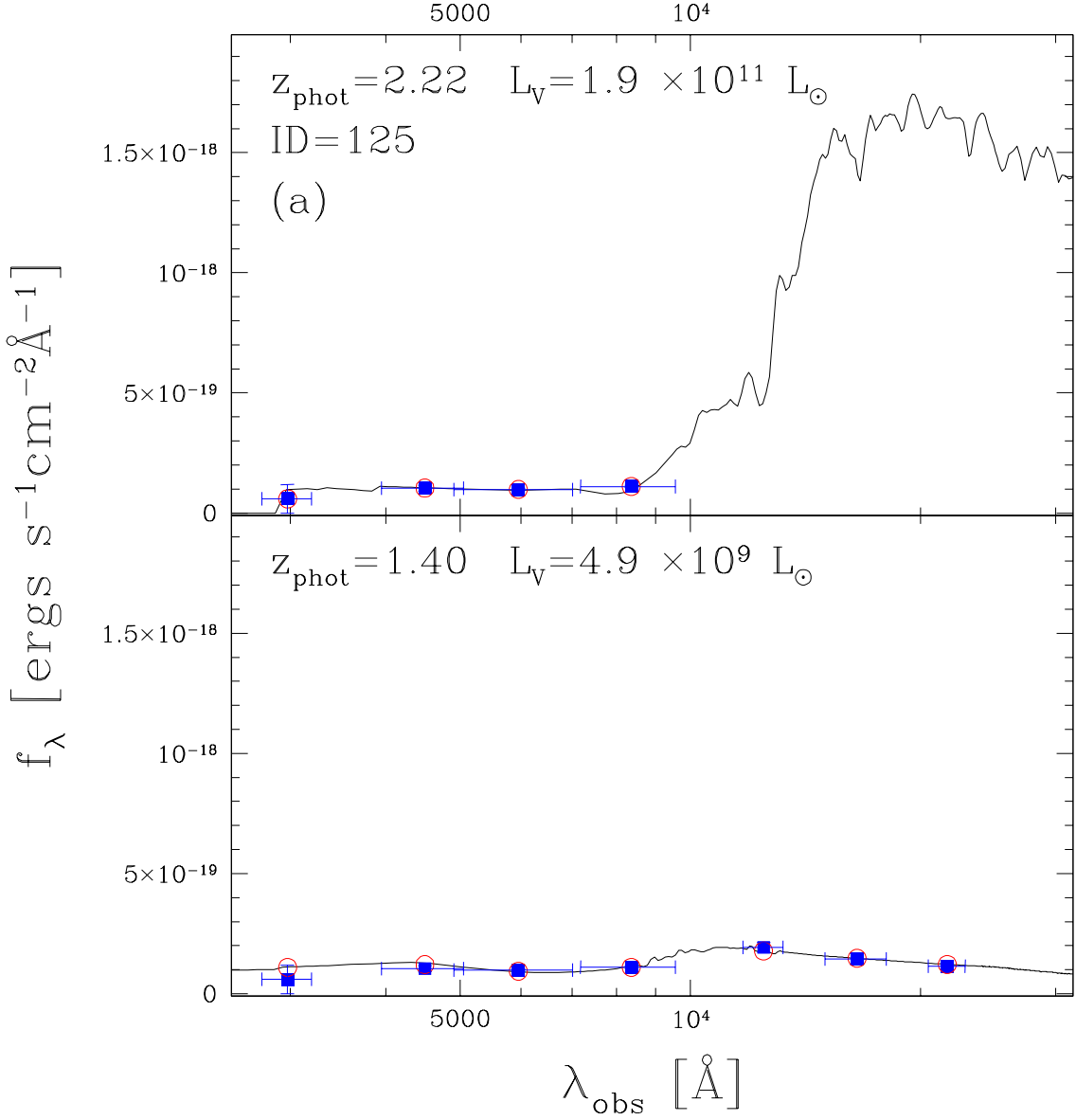}}
\caption{
  Two examples of how the inclusion of near infrared data helps to
  measure the correct \zp. Obviously, the inferred \lrest~is strongly
  coupled to \zp.  The top panels for each object contain the fit
  using only data from the four optical HST filters.  The bottom
  panels contain the fit using all seven bands.  The solid points are
  the data and the empty points are the model fluxes.}
\label{ircomp}
\end{figure}

To demonstrate the effect of the inclusion of deep NIR data in the
redshift range $1.5<z<2$, we show in Figure~\ref{ircomp} two galaxies
fit with and without the NIR information.  Even where the $V-I$ color
is well constrained, and hence the possible redshifts severely
limited, the NIR data can fix the break position.

The three highest redshift objects in our sample (objects 542, 424,
and 45) have \zp=3.86, 4.82, and 5.34 and \Ksabtot=22.75, 23.29, and
23.16 respectively.  Object 542 has 68\% redshift confidence limits of
\zp=0.42-3.88.  In general, while the observed SED of object 424 is
fit well, there is flux blueward of the predicted 912\ang~break
position.  The high redshift is chosen by the technique because the
red H-\Ks~color indicates the presence of a rest-frame optical break.
No Monte-Carlo realizations end up in a secondary minimum, but when
fit using only the optical data, a redshift of 1.1 is found.  Object
45 has a poor fit in the NIR, and has a redshift of 1.34 when fit with
only the optical data.  We do not consider these objects in any of our
analyses.

\subsection{The Redshift Distribution}

\begin{figure}
\scalebox{0.8}{\includegraphics{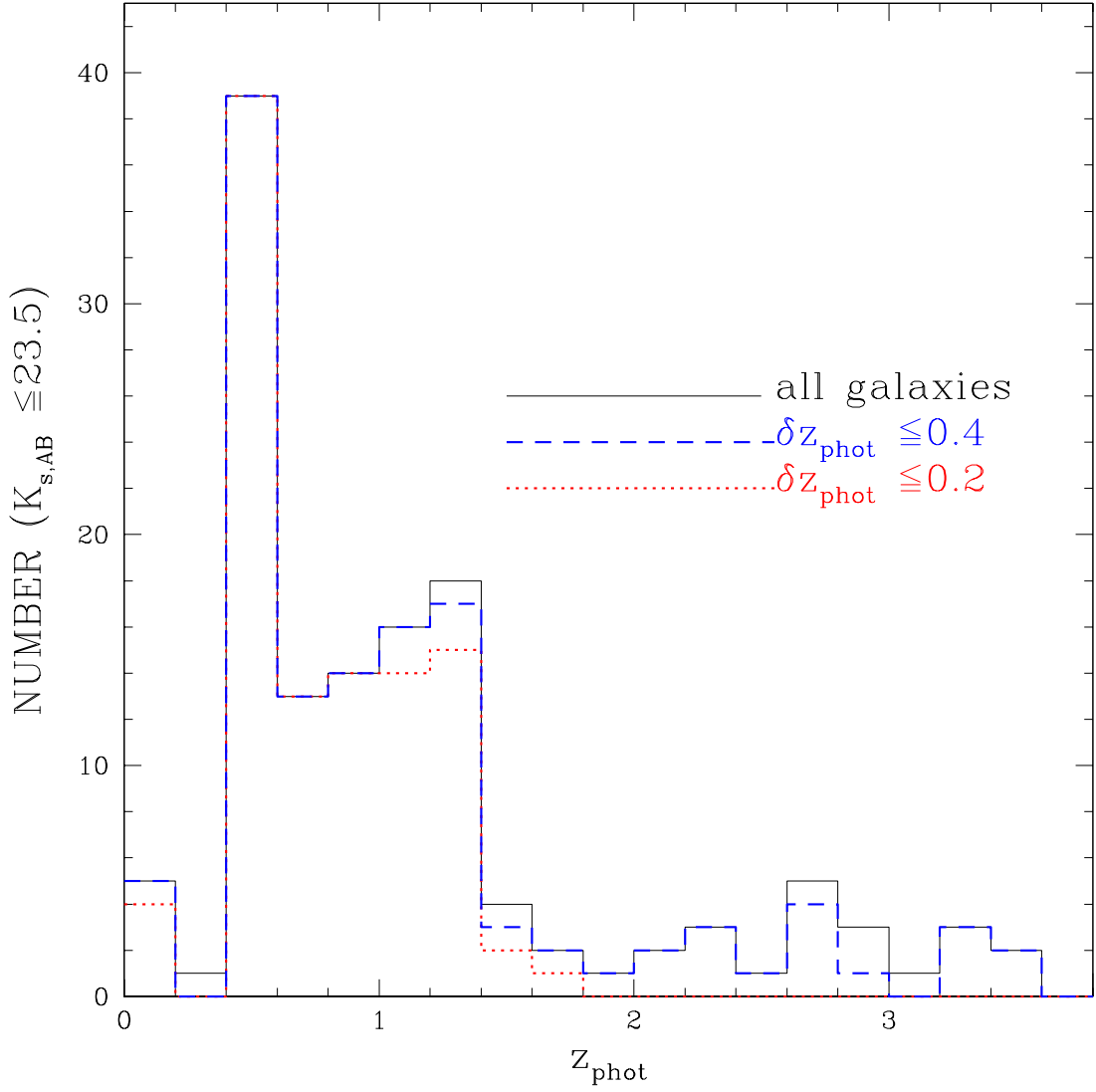}}
\caption{
  The redshift histogram of all 132 objects in our catalog with
  reliable redshifts (solid line).  The two other histograms show the
  redshift distributions for all objects with \dzp~$\leq~0.4$ (dashed
  line) and all objects with \dzp~$\leq~0.2$ (dotted line) where the
  photometric redshift errors are the combination of those calculated
  using our Monte Carlo technique with the systematic errors
  determined from the HDF-N.}
\label{zhist}
\end{figure}

\begin{figure}
\begin{center}
\scalebox{0.8}{\includegraphics{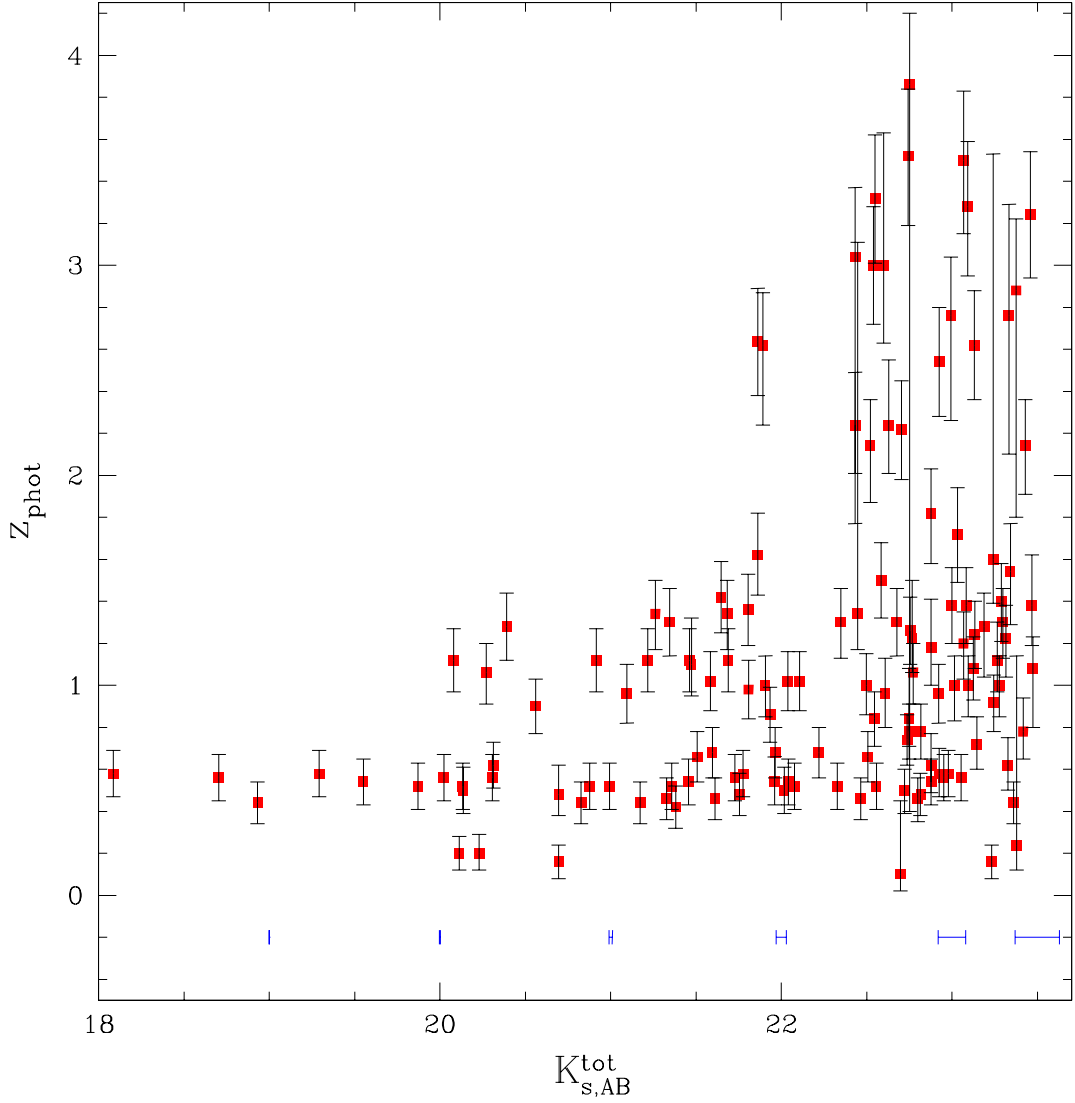}}
\end{center}
\caption{
  The \Ksabtot~magnitude of our objects vs. \zp.  The photometric
  redshift errors are a combination of those calculated using our
  Monte Carlo technique and the systematic errors calculated from
  agreement with spectroscopic redshifts in the HDF-N.  At the bottom
  of the graph, we show the typical photometry errors of objects of
  different magnitude.}
\label{magz}
\end{figure}

In Figure~\ref{zhist} we show the histogram of the photometric
redshifts listed in Table 3.  The three sets of lines represent
galaxies with different photometric redshift precision.  This figure
also reveals structure in the redshift histogram with a sharp peak at
\zp$\approx 0.5$ and a broad enhancement at $1\leq$\zp$\leq 1.4$.  The
redshift peak at $z\approx 0.5$ was first noticed by G01 from AAT
spectroscopic redshifts taken over a larger field centered on the
HDF-S.  To examine the luminosity distribution of galaxies in these
enhancements, we plot \zp~vs. \Ksabtot~in Figure~\ref{magz}, revealing
that they are prominent in very bright galaxies, \Ksabtot$<$21.5.
These strong features in our redshift distribution are also seen in a
\Ksabtot$\leq 23.5$ subsample of the HDF-S data from Fontana et al.
(2000).  HDF-N contains several peaks, but they are not as strong as
the features in the HDF-S (Cohen et al. 1996).

\begin{figure}
\begin{center}
\scalebox{0.8}{\includegraphics{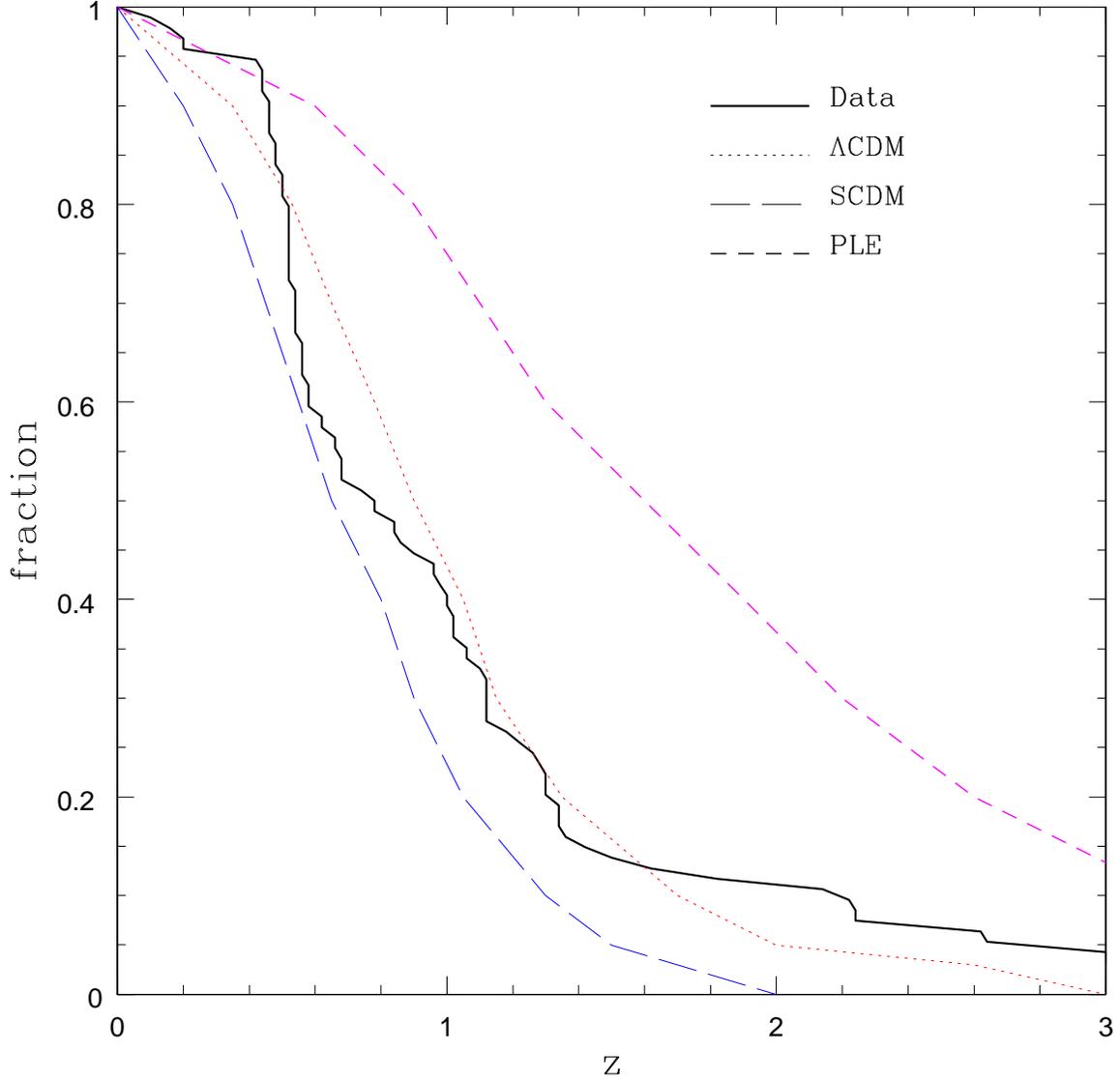}}
\end{center}
\caption{
  The cumulative redshift histogram for the 95 galaxies in our sample
  with \Ks$^{vega}<21$ as indicated by the solid curve.  The other
  curves are semi-analytical model predictions from Fontana \etal
  (1999) for an SCDM ($\Omega_m=1.0,~\Lambda=0.0,~h=0.5$; long dashed
  curve), $\Lambda$CDM ($\Omega_m=0.3,~\Lambda=0.7,~h=0.6$; dotted
  curve), and PLE model. The data are generally consistent with
  hierarchical models of formation while the PLE model significantly
  overpredicts the number of bright galaxies at at high redshift.  }
\label{z_cumu}
\end{figure}

We can use the overall redshift distribution of galaxies in our sample
to test the predictions of theoretical models of galaxy formation.  In
Figure~\ref{z_cumu} we directly compare our cumulative redshift
distribution for galaxies with \Ksvega$<21$ to the theoretical
predictions for SCDM ($\Omega_m=1.0,~\Lambda=0.0,~h=0.5$),
$\Lambda$CMD ($\Omega_m=0.3,~\Lambda=0.7,~h=0.6$) and Pure Luminosity
Evolution (PLE) models calculated by F99 following slightly modified
versions of the KC98 prescriptions.  At almost all redshifts, SCDM
underpredicts the fraction of galaxies which lie at high redshifts,
while the $\Lambda$CDM model provides a much better description of the
data.  Both CDM models reproduce the median redshift of the data
($z\sim~0.8$) reasonably well.  The difference between the CDM models
can be understood because galaxy formation occurs at higher redshift
in a $\Lambda$ dominated universe.  It is also interesting to note
that the PLE models severely overpredict the abundance of bright
galaxies at all redshifts.  Our data has a low ($\lesssim 1\%$) K-S
probability of being drawn from any of the models.  This is likely due
to the clustering of galaxies in our small volume as the CDM models
reproduce the general trends well.  We note however that the models do
not take into account any of the observational biases and
incompleteness that may occur for IR selected galaxies.  NIR selection
is generally thought to be less prone to extinction effects and less
dependent on the current SFR than optical selection.  However surface
brightness dimming and the bright IR sky can limit detection
efficiency for extended objects.

We now compare our results directly with those of F99 and the SUNY
group.  F99 claims that in a \Ksvega$<21$ sample, only 2\% of the
galaxies lie at \zp$\geq2$ in the HDF-S and 6\% in the NTT Deep Field.
In contrast, we find in our data that 12\% of the galaxies with
\Ksvega$<21$ lie at \zp$\geq2$.  Using a \Ksvega$<21$ subsample of the
SUNY Stonybrook HDF-S photometric redshift catalog we find that the
fraction of galaxies lying at \zp$\geq2$ is identical to ours.  The
differences between us and F99 are not due to small sample selection
issues.  There are 5 galaxies with \Ksvega$<21$ which F99 place at
$1.5 < z < 2$ but which we find at $2 < z < 3$.  The exact differences
between the high-redshift fractions measured by different photometric
redshift techniques can depend rather sensitively on the redshift
threshold used to discriminate between ``high'' and ``low'' redshift
galaxies.  For example, although there is disagreement on the fraction
of galaxies at \zp$\geq2$ both F99 and we are in agreement about the
fraction of the \Ksvega$<21$ galaxies ($\sim 14-15\%$) in the HDF-S
which lie at \zp$\geq1.5$.  These discrepancies will be eventually
resolved with extensive spectroscopy in the NIR and the blue optical.

\subsection{Rest-Frame Luminosities}
\label{lumin}

\begin{figure}
\scalebox{0.8}{\includegraphics{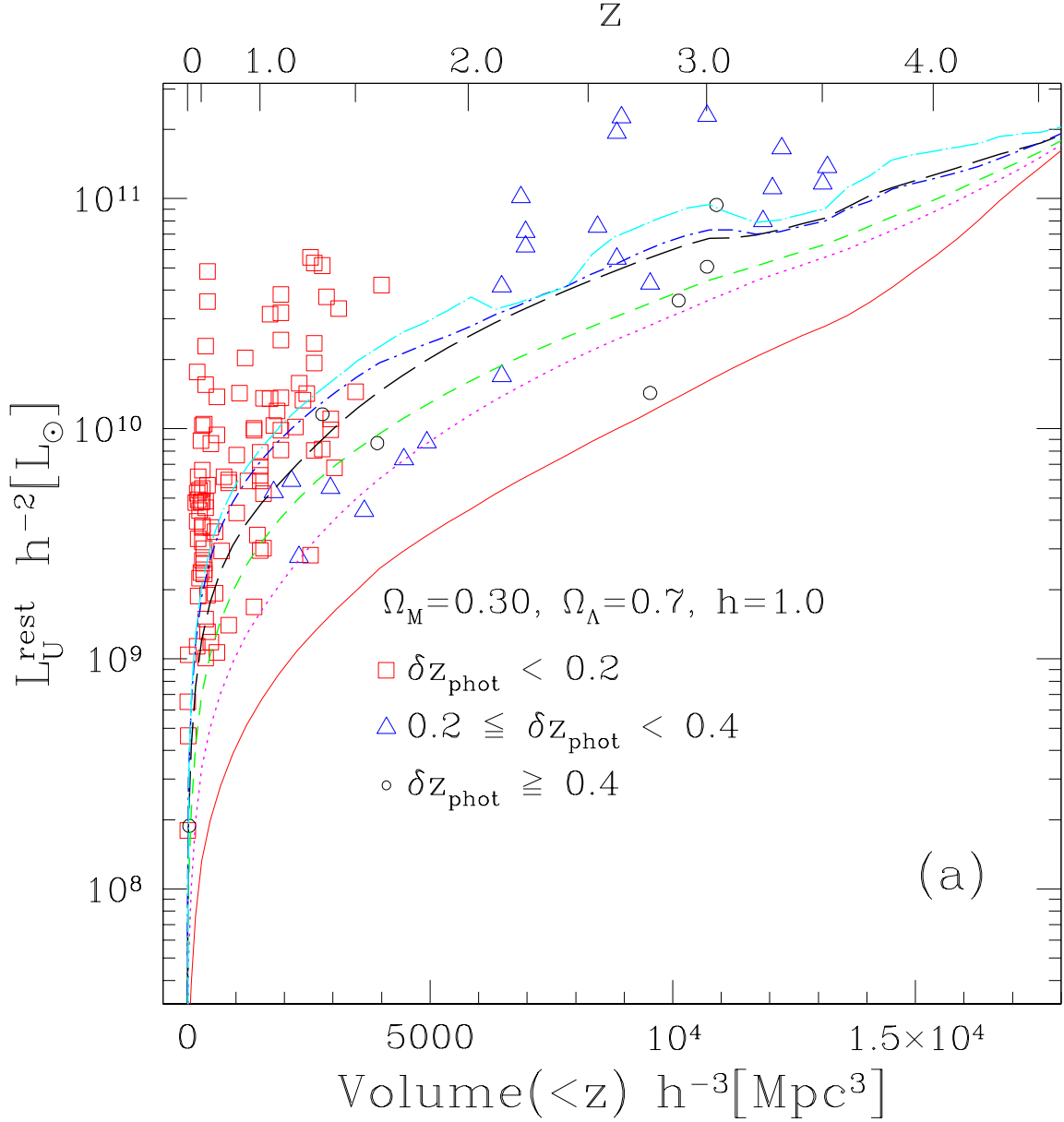}}
\caption{
  The distribution of rest-frame U, B, and V-band luminosities as a
  function of enclosed co-moving volume and \zp~is shown in figures a,
  b, and c respectively.  We show all 132 galaxies with
  \Ksab~$\leq$~23.5 and reliable redshift estimates.  Note the large
  number of galaxies at \zp$\geq 2$ with \lrest~$\geq 5\times
  10^{10}L_{\odot}$.  The tracks represent the values of \lrest~for
  each our six template spectra normalized at each redshift to
  \Ksab$=23.5$.  The large star in \textbf{b)} indicates the value of
  \lstarb~from local surveys.  The specific tracks correspond to the E
  (solid), Sbc (dot), Scd (short dash), Irr (long dash), SB1
  (dot--short dash), and SB2 (dot--long dash) templates.}
\label{lumz}
\end{figure}

Our long wavelength baseline allows us to observe a given rest-frame
wavelength over a large range in redshift.  From the best-fit SED at
the best-fit redshift we measured the rest-frame luminosity in the U,
B, and V bands for our galaxies and plot this as a function of
enclosed volume and redshift in Figure~\ref{lumz}.  As reference to
solar values, we take $2.73\times 10^{29},~5.10\times 10^{29}$, and
$4.94\times 10^{29}~{\rm ergs~s^{-1}\AA^{-1}}$ for L$_{\odot}^U$,
L$_{\odot}^B$, and L$_{\odot}^V$ respectively (assuming $M_U=+5.66,
M_B=+5.47$, and $M_V=+4.82$ in Johnson magnitudes; Cox 2000).  Using
the distribution of \lrest~values measured over \dzp, we calculate an
errorbar in \lrest~for each galaxy.  While we differentiate points in
Figure~\ref{lumz} based on their values of \dzp, the errors in
\lrest~are tightly coupled with the values of \dzp~and so are not
presented on this plot.  This coupling is demonstrated by the two
cases in Figure~\ref{ircomp} where the main uncertainty in
\lrest~stems from the uncertainty in \zp, not from the specific values
of the NIR data.  All values of \lrest~and their associated
uncertainties are presented in Table 3.

Because our fluxes are measured in uncorrected 2\farcs0 apertures, we
may be missing flux for the larger galaxies.  Therefore, we correct
all values of \lrest~by the ratio (in the \Ks-band) of the SExtractor
total flux to the 2\farcs0 aperture flux.  The median correction
factor is 1.05 with $68\%$ confidence limits of 0.97 and 1.25.  The
largest correction is by a factor of 1.72.  To quantitatively asses
the goodness of our SED fits we compared the luminosities derived from
the best-fit SED to the luminosities derived from a linear
interpolation between the observed filters shifted to the desired
redshift, and found the RMS differences to be $\lesssim10\%$ in all
bands.

Perhaps the most interesting feature of Figure~\ref{lumz} is the
presence of intrinsically luminous galaxies (\lrest$\geq 5\times
10^{10}h^{-2}L_{\odot}$) in all passbands at high redshifts.  The
apparent lack of low luminosity galaxies at high redshift in
Figure~\ref{lumz} merely reflects our \Ks~magnitude limit translated
to a rest-frame luminosity limit.  Also apparent in Figure~\ref{lumz},
at $z>1$, is the increasing range in \lrest~toward shorter rest-frame
wavelengths.  This is due to our magnitude limit in \Ks, combined with
the variation in intrinsic galaxy colors.  We demonstrate this by
showing the \lrest-z tracks of our 6 galaxy templates normalized to
\Ksab$=23.5$.

We use the local B-band luminosity function to estimate the evolution
in the bright high-redshift galaxies.  We find 9 galaxies with \lb$
\geq 5\times 10^{10}h^{-2}L_{\odot ,B}$ which lie in a volume of
$7.29\times 10^3 h^{-3} Mpc^3$ between $2\leq z\leq3.5$.  We should be
at least $50\%$ complete for all galaxy types over this redshift and
luminosity range.  The number of galaxies at the bright end of the
luminosity function is especially sensitive to variations in
\lstar~and we try to measure evolution in the luminosity function by
holding $\alpha$ and \phistar~constant and changing \lstar~to match
the observed counts.  We use the local luminosity functions derived
from the Sloan Digital Sky Survey (SDSS; Blanton \etal 2001) and the
2dF Galaxy Redshift Survey (2dFGRS; Folkes \etal 1999) to predict the
number of galaxies expected in this volume.  The 2dFGRS luminosity
function is in $b_j$ magnitudes and Blanton \etal (2001) provide a
conversion of their SDSS luminosity function to this system.  With $B
= b_j + 0.2$ for a typical galaxy color of $(B-V)\approx 0.6$, the
SDSS luminosity function then gives \lstarb$ = 9.7 \times
10^9h^{-2}L_{\odot,B}$, \phistar$=2.69\times 10^{-2} h^{3} Mpc^{-3}$,
and $\alpha = -1.22$ while the 2dFGRS gives \lstarb$ = 1.0 \times
10^{10}h^{-2} L_{\odot,B}$, \phistar$=1.69\times 10^{-2} h^{3}
Mpc^{-3}$, and $\alpha = -1.28$.  The predicted numbers of galaxies in
this volume are $\approx0.1$ for both the SDSS and 2dFGRS luminosity
functions. If \lstarb~is increased by a factor of 2.7 or 3.2 for the
SDSS and 2dFGRS luminosity functions, respectively, then 9 galaxies
are predicted.  Because of the small co-moving volumes enclosed in
this redshift range, these numbers may not be indicative of the galaxy
population as a whole.  Furthermore, random errors in the photometric
redshifts will tend to produce a bias in the derived luminosities, as
the luminosity function declines very steeply towards higher
luminosities, and the smoothing will increase the number of observed
very luminous galaxies.  We estimate this effect by convolving the
Schechter function with a Gaussian of width 0.3 magnitudes
characteristic of our errors.  As a result, the required increase in
\lstar~decreases to 2.4-2.9 with respect to locally determined values.
It is clear that spectroscopic confirmation of the photometric
redshifts of these bright galaxies is desirable.

Another striking feature is the lack of galaxies with \lv$\gtrsim
1.4\times 10^{10}h^{-2}L_{\odot}$ and $1.5< z < 2$.  Given the
observed redshift structure in our field, this may simply be due to
clustering.  It is interesting however to note that Dickinson (2001b)
found a similar paucity of intrinsically luminous galaxies at
$1.4<z<2$ in the HDF-N.  The photometric redshifts in this regime are
particularly uncertain however, as spectroscopic redshifts are rarely
available.  The derived \zp~between $1.5 < z < 2.5$ is very sensitive
to the U-band photometry, as the Lyman break moves into the U-band.
We tested how \zp~changes if the U-band data is omitted.  The largest
changes occur for galaxies with $2 < z < 2.5$, and their newly derived
\zp~are systematically lower.  This suggests that \zp~might be biased
if the bluest band falls just above the rest-frame Lyman break.

\section{Summary and Conclusions}

We have presented the initial results from the \textbf{F}aint
\textbf{I}nfra-\textbf{R}ed \textbf{E}xtragalactic \textbf{S}urvey
(FIRES) obtained with ISAAC at the VLT.  We assembled a \Ks-band
selected catalog of galaxies in the HDF-S from the deepest NIR data
taken of this field.  Our catalog consists of 136 galaxies with
\Ksab$\leq23.5$ and photometry in seven bands from $0.3\mu$m to
$2.2\mu$m.  Our unique combination of ultra-deep optical data from HST
with our deep NIR data allows us to sample the rest-frame V-band in
galaxies for $z\leq3$ and to select galaxies in a way less dependent
on the current SFR than the rest-frame UV.

To interpret these data, we have developed a new photometric redshift
algorithm which models the galaxy colors with a linear combination of
empirical templates and in so doing, makes minimal {\it a priori}
assumptions about the galaxies' SFH.  Testing our method on galaxies
with spectroscopic redshifts from the HDF-N and HDF-S, we find that
our technique is precise and robust for all \zs$<6$ having a mean
$\Delta z\approx 0.10$ for $z \leq 1.5$ and $\Delta z\approx 0.44$ for
$z > 1.5$ with catastrophic errors in $\lesssim3\%$ of the sample.
The results from the HDF-S also confirm that our photometry is
adequate for good \zp~estimates.  We find that in almost all
cases that our best-fit SED matches the observed fluxes well.

We developed a Monte-Carlo code to estimate the uncertainty in \zp~
arising from the flux errors.  In agreement with previous work by
other groups, we found that the uncertainty in \zp~is dominated at the
faint end by photometric uncertainty, and at the bright end by
template mismatch.  For bright galaxies, where spectroscopic redshifts
are available, the uncertainty in \zp~is severely underestimated when
it is derived solely from the flux uncertainties although large values
of \dzm~can help identify catastrophic errors in \zp.  To provide
realistic, individual estimates on the accuracy each galaxy's \zp~we
added our Monte-Carlo errors in quadrature with the mean disagreement
with \zs~as measured from the HDF-N and also flag galaxies with
secondary minima in their $\chi^2(z)$ profiles.

Although the redshift is primarily constrained by the high
signal-to-noise HST optical data, the deep NIR data can break
degeneracies between different template combinations at different
redshifts, which have identical $V-I$ colors.  While the NIR data
greatly improves the redshift estimation at $z<1.5$, it can actually
worsen the \zp~estimate at high redshifts by causing the
mis-identification of a Lyman break as a rest-frame optical break.
The effect of the NIR should become increasingly important when the
signal-to-noise is dramatically improved, such as in the very deep
exposures planned for FIRES.  By fixing the position of rest-frame
optical breaks at $z>1$, our NIR data also allows us to probe the
redshift distribution of all galaxy types at these epochs.  We use our
photometric redshift technique to estimate \zp~and its accompanying
uncertainty for our entire \Ks-band selected sample.

Applying these techniques, we have found a sharp peak in the redshift
distribution at $z\approx 0.5$ and an broad peak at $1\leq$\zp$\leq
1.4$.  The $z\approx 0.5$ spike was first noticed by G01 using
spectroscopic redshifts obtained with the AAT.

To compare our redshift distribution with the predictions of
hierarchical galaxy formation models, we measured the fraction of
galaxies at $z>2$ in a \Ksvega$<21$ sample to be $12\%$.  We find
that this fraction is much greater than that predicted by KC98 and F99
for a CDM universe with $\Omega_m = 1$ although it is in better
agreement with a $\Lambda$CDM model.  At all redshifts we find far
fewer bright galaxies than predicted by PLE models.  We also find
however, that different groups working with similar datasets find
different fractions of galaxies above a certain redshift threshold.
This disagreement stems from differences in \zp~determinations between
groups.

Taking advantage of our extended wavelength coverage, we measure the
rest-frame luminosity \lrest~in the U, B, and V bands for the galaxies
in our sample, regardless of their redshift.  Many high-redshift
galaxies have \lrest$\geq 5\times 10^{10}~h^{-2}L_{\odot}$ in all
bands, however we find a paucity of galaxies with \lv$\geq 1.4\times
10^{10}~h^{-2}L_{\odot}$ between $1.5<z<2$.  A similar deficit in the
redshift distribution of intrinsically luminous galaxies was noted by
Dickinson (2001b) using NICMOS data on the HDF-N.  However, the
photometric redshifts in this regime are uncertain and spectroscopic
confirmation of this deficit is required.  At higher redshifts the
densities increase and we find 9 galaxies with \lb$ \geq 5\times
10^{10}h^{-2}L_{\odot ,B}$ which lie between $2\leq z\leq3.5$.  These
numbers can be accounted for if \lstar~in the B-band increases by a
factor of 2.7-3.2 with respect to SDSS and 2dFGRS values.  When
accounting for uncertainties in the rest-frame luminosity, the
required increase is 2.4-2.9.  The redshifts and nature of these
intrinsically bright galaxies at high-z needs to be verified with
spectroscopic follow-up.

It is tempting to associate the increase in the number density of
bright galaxies at $z<1.5$ compared to $1.5<z<2$ with the onset of
disk formation.  Spectroscopic studies of larger volumes are necessary
to rule out that cosmic variance, or uncertainties in the photometric
redshifts dominate this effect.

\acknowledgments

GR would like to thank Marc Sarzi, Thilo Kranz, and Nicolas Cretton
for many useful discussions.  We would like to thank the ESO staff for
their assistance and their efforts in taking these data and making
them available to us.  GR thanks Leiden University for its hospitality
during several working trips.  MF and GR thank MPIA for travel support.

\clearpage

\begin{figure}
\begin{center}
\end{center}
\caption{
  The reduced \Ks-band image.  All 136 objects in the final catalog
  are marked, and the numbers are the ID numbers in the catalog shown
  in Table 1.  The outline of the WFPC2 field of the HDF-S is shown.
}
\label{image}
\end{figure}

\begin{figure}
\scalebox{0.8}{\includegraphics{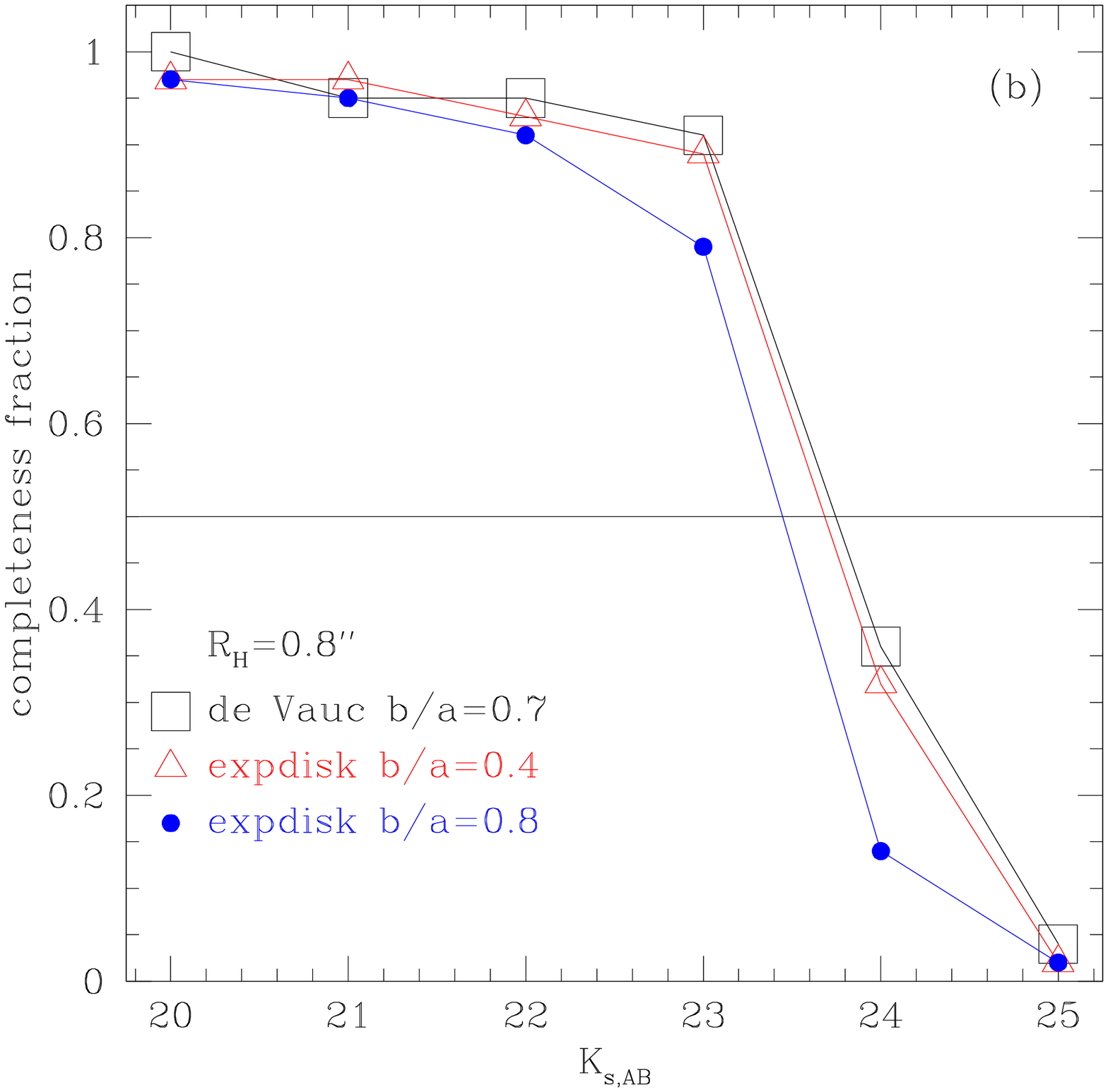}}
\end{figure}

\begin{figure}
\scalebox{0.8}{\includegraphics{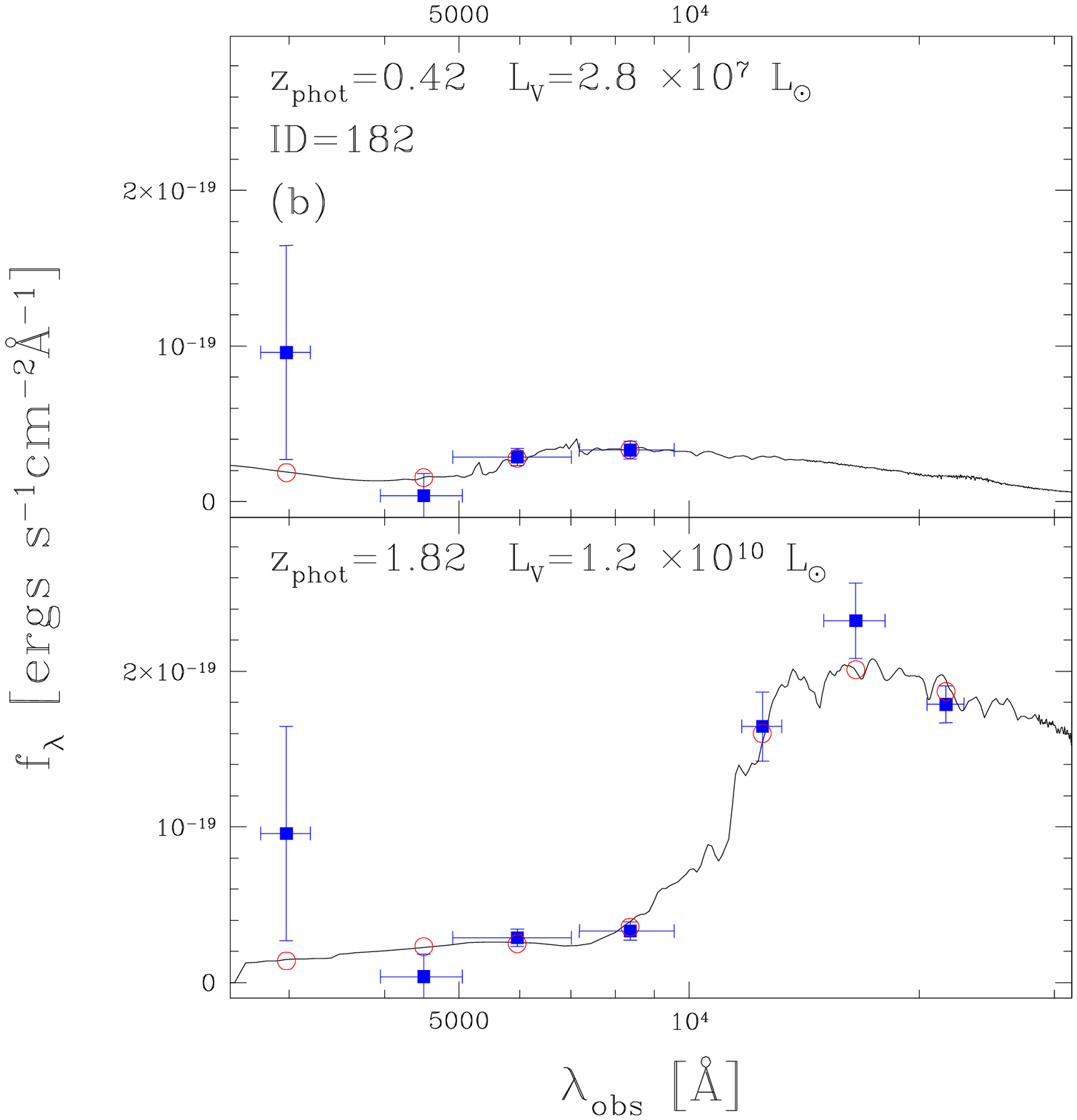}}
\end{figure}

\begin{figure}
\scalebox{0.8}{\includegraphics{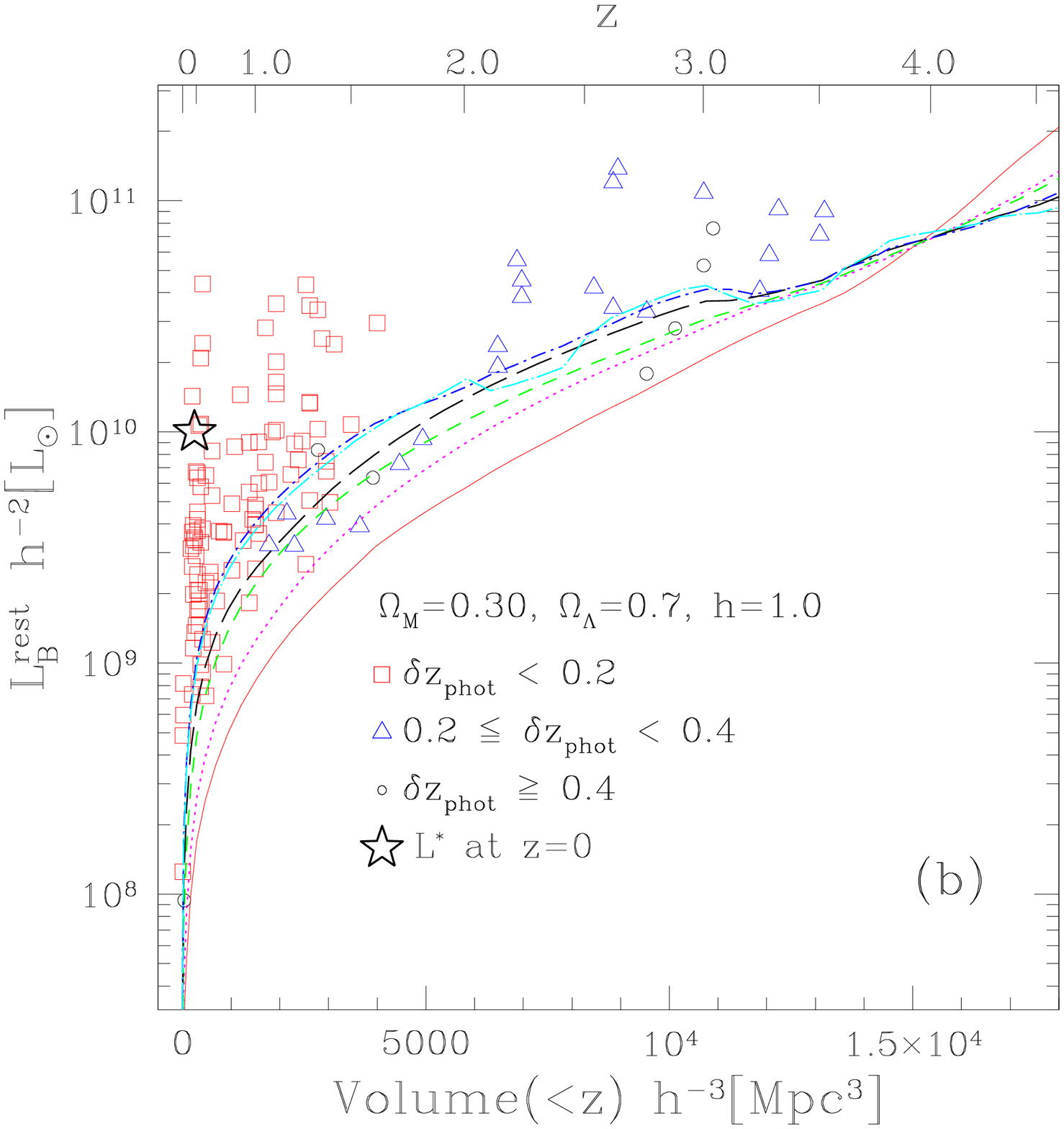}}
\end{figure}
 
\begin{figure}
\scalebox{0.8}{\includegraphics{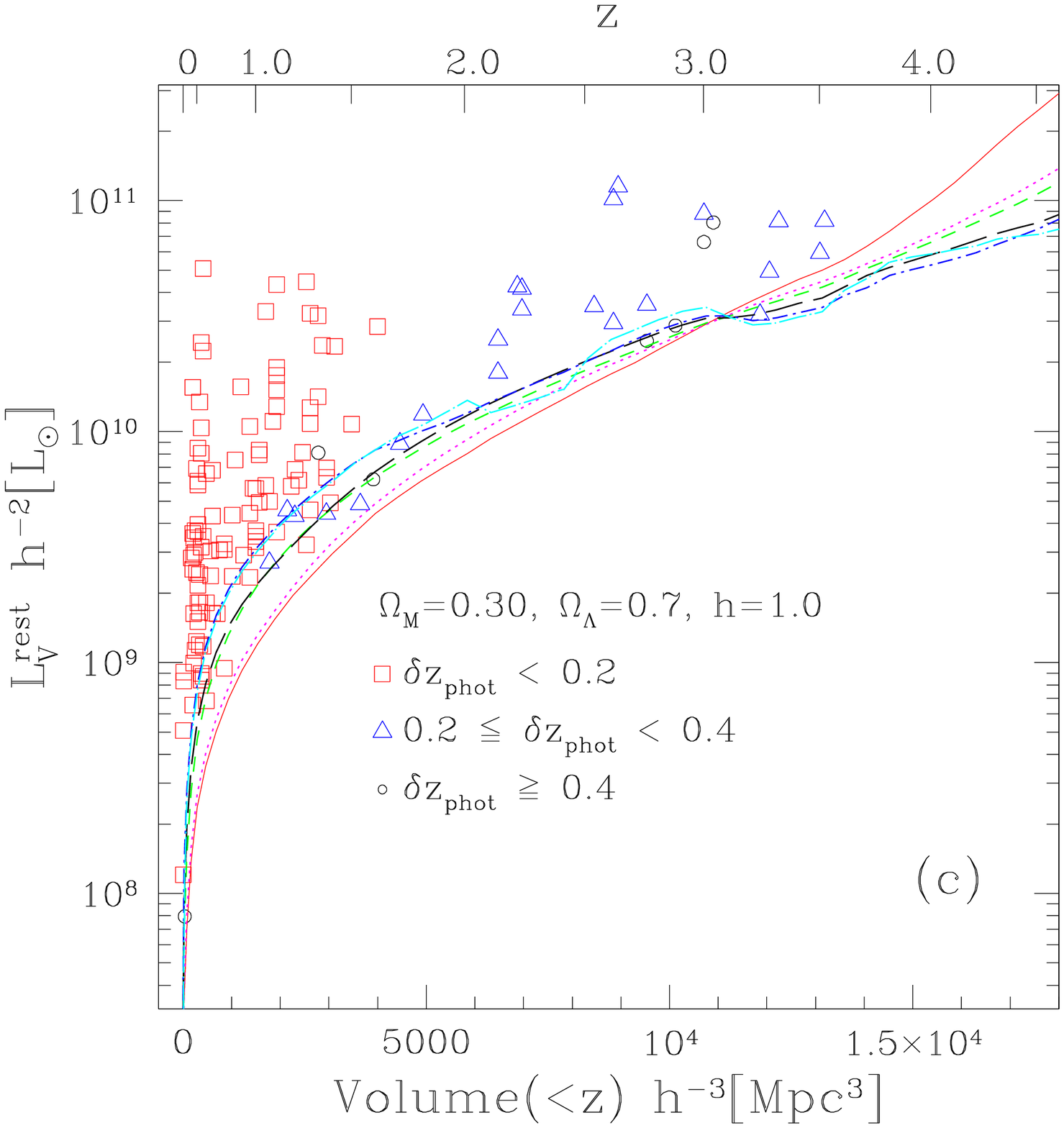}}
\end{figure}
 
\clearpage

\begin{deluxetable}{lllllllll}
\rotate
\tablewidth{0pt}
\tabletypesize{\footnotesize}
\tablecaption{Photometric~Catalog}
\tablehead{\colhead{ID} & \colhead{F300W\tablenotemark{a} } & \colhead{F450W\tablenotemark{a} } & \colhead{F606W\tablenotemark{a} } & \colhead{F814W\tablenotemark{a} } & \colhead{${\rm J_s}$\tablenotemark{a} } & \colhead{H\tablenotemark{a} } & \colhead{${\rm K_s}$\tablenotemark{a} } & \colhead{${\rm K_s^{tot~b}}$
}\\
}
\startdata
${\rm~HDFS1-30}$ & $17.5\pm2.0$ & $62.4\pm1.0$ & $81.7\pm0.7$ & $140.4\pm1.2$ & $465.8\pm15.2$ & $528.4\pm29.2$ & $629.7\pm24.8$ & $700.4\pm39.6$ \\ 
${\rm~HDFS1-33}$ & $19.2\pm1.9$ & $37.6\pm1.0$ & $45.2\pm0.7$ & $93.9\pm1.2$ & $128.3\pm15.1$ & $195.7\pm29.0$ & $184.7\pm24.7$ & $186.5\pm24.5$ \\ 
${\rm~HDFS1-31}$ & $6.9\pm1.9$ & $11.7\pm1.0$ & $21.9\pm0.6$ & $41.1\pm1.2$ & $71.1\pm14.1$ & $119.7\pm27.4$ & $184.1\pm23.1$ & $261.5\pm38.4$ \\ 
${\rm~HDFS1-36}$ & $0.3\pm2.0$ & $34.9\pm1.0$ & $73.8\pm0.7$ & $90.5\pm1.2$ & $119.4\pm13.5$ & $275.2\pm26.2$ & $337.0\pm22.0$ & $353.4\pm28.0$ \\ 
${\rm~HDFS1-37}$ & $4.6\pm2.0$ & $1.0\pm1.0$ & $3.5\pm0.7$ & $6.2\pm1.2$ & $11.3\pm14.7$ & $119.1\pm28.3$ & $281.7\pm24.0$ & $337.1\pm37.1$ \\ 
${\rm~HDFS1-45}$ & $1.9\pm2.0$ & $1.4\pm1.0$ & $9.6\pm0.7$ & $71.4\pm1.2$ & $228.4\pm14.7$ & $225.4\pm28.4$ & $216.9\pm24.1$ & $201.1\pm22.4$ \\ 
${\rm~HDFS1-50}$ & $0.3\pm1.9$ & $2.0\pm1.0$ & $5.4\pm0.7$ & $18.3\pm1.2$ & $132.0\pm12.4$ & $247.2\pm23.6$ & $294.2\pm19.9$ & $289.0\pm20.2$ \\ 
${\rm~HDFS1-52}$ & $30.9\pm1.9$ & $53.8\pm1.0$ & $60.1\pm0.6$ & $85.1\pm1.2$ & $139.0\pm13.4$ & $151.8\pm25.9$ & $146.9\pm21.8$ & $174.4\pm26.5$ \\ 
${\rm~HDFS1-54}$ & $16.2\pm1.9$ & $23.5\pm1.0$ & $31.6\pm0.6$ & $52.4\pm1.2$ & $98.0\pm13.0$ & $127.7\pm24.9$ & $159.0\pm21.1$ & $150.8\pm28.7$ \\ 
${\rm~HDFS1-62}$ & $7.7\pm1.9$ & $28.6\pm0.9$ & $37.3\pm0.6$ & $75.3\pm1.2$ & $131.4\pm11.3$ & $140.9\pm21.6$ & $220.0\pm18.2$ & $230.4\pm29.9$ \\ 
${\rm~HDFS1-58}$ & $4.3\pm1.9$ & $4.3\pm1.0$ & $17.5\pm0.6$ & $75.1\pm1.2$ & $388.1\pm12.1$ & $576.8\pm23.1$ & $782.9\pm19.5$ & $858.8\pm29.7$ \\ 
${\rm~HDFS1-63}$ & $62.9\pm1.9$ & $124.9\pm0.9$ & $287.9\pm0.6$ & $491.4\pm1.2$ & $823.6\pm11.5$ & $986.8\pm22.0$ & $1179.9\pm18.5$ & $1255.0\pm26.3$ \\ 
${\rm~HDFS1-69}$ & $22.3\pm1.9$ & $32.6\pm0.9$ & $46.6\pm0.6$ & $92.8\pm1.2$ & $193.9\pm11.5$ & $248.7\pm22.0$ & $308.5\pm18.5$ & $294.5\pm18.2$ \\ 
${\rm~HDFS1-74}$ & $8.9\pm1.9$ & $22.0\pm0.9$ & $51.7\pm0.6$ & $166.1\pm1.2$ & $536.9\pm11.5$ & $858.4\pm22.0$ & $1248.0\pm18.5$ & $1349.0\pm26.6$ \\ 
${\rm~HDFS1-79}$ & $25.2\pm1.9$ & $83.7\pm0.9$ & $99.7\pm0.6$ & $120.1\pm1.2$ & $211.8\pm11.5$ & $236.9\pm22.0$ & $249.8\pm18.5$ & $306.6\pm28.0$ \\ 
${\rm~HDFS1-80}$ & $-0.3\pm1.9$ & $23.6\pm0.9$ & $44.2\pm0.6$ & $57.0\pm1.2$ & $80.2\pm11.5$ & $63.5\pm22.0$ & $165.0\pm18.5$ & $153.0\pm17.2$ \\ 
${\rm~HDFS1-83}$ & $103.7\pm1.9$ & $163.8\pm0.9$ & $338.9\pm0.6$ & $534.6\pm1.2$ & $739.3\pm11.5$ & $784.4\pm22.0$ & $896.1\pm18.5$ & $1087.8\pm30.1$ \\ 
${\rm~HDFS1-86}$ & $14.4\pm3.1$ & $93.2\pm1.2$ & $177.8\pm0.9$ & $256.5\pm1.7$ & $308.0\pm13.0$ & $312.9\pm24.5$ & $184.1\pm20.6$ & $188.3\pm22.1$ \\ 
${\rm~HDFS1-87}$ & $2.0\pm1.9$ & $5.4\pm1.0$ & $12.3\pm0.6$ & $19.4\pm1.2$ & $83.6\pm11.5$ & $79.6\pm22.1$ & $196.8\pm18.6$ & $187.0\pm18.0$ \\ 
${\rm~HDFS1-92}$ & $8.3\pm2.0$ & $16.7\pm1.0$ & $24.7\pm0.7$ & $41.4\pm1.2$ & $134.6\pm11.5$ & $216.8\pm22.0$ & $231.2\pm18.5$ & $216.0\pm17.2$ \\ 
${\rm~HDFS1-98}$ & $2.3\pm2.0$ & $38.6\pm1.0$ & $221.7\pm0.6$ & $846.6\pm1.2$ & $2035.4\pm11.5$ & $2785.0\pm22.0$ & $3513.2\pm18.5$ & $3620.1\pm26.7$ \\ 
${\rm~HDFS1-105}$ & $-3.6\pm1.9$ & $3.1\pm0.9$ & $2.1\pm0.6$ & $7.5\pm1.2$ & $45.4\pm11.5$ & $235.4\pm22.0$ & $368.2\pm18.6$ & $363.0\pm19.5$ \\ 
${\rm~HDFS1-107}$ & $34.3\pm1.9$ & $48.0\pm0.9$ & $60.6\pm0.6$ & $101.8\pm1.2$ & $156.7\pm11.5$ & $232.0\pm22.0$ & $195.1\pm18.5$ & $181.0\pm17.2$ \\ 
${\rm~HDFS1-99}$ & $16.1\pm1.8$ & $27.7\pm0.9$ & $44.1\pm0.6$ & $89.5\pm1.1$ & $145.3\pm10.6$ & $152.6\pm20.3$ & $180.7\pm17.1$ & $203.9\pm26.5$ \\ 
${\rm~HDFS1-119}$ & $35.4\pm1.9$ & $50.4\pm0.9$ & $75.9\pm0.6$ & $166.4\pm1.2$ & $304.0\pm11.5$ & $305.8\pm22.0$ & $339.7\pm18.5$ & $354.3\pm21.8$ \\ 
${\rm~HDFS1-111}$ & $7.0\pm1.9$ & $50.8\pm0.9$ & $251.3\pm0.6$ & $802.5\pm1.2$ & $1931.8\pm11.5$ & $2662.9\pm22.0$ & $3228.2\pm18.5$ & $3267.8\pm26.2$ \\ 
${\rm~HDFS1-112}$ & $15.2\pm1.9$ & $43.8\pm0.9$ & $52.8\pm0.6$ & $64.5\pm1.2$ & $131.9\pm11.4$ & $183.6\pm22.0$ & $153.4\pm18.5$ & $157.2\pm30.2$ \\ 
${\rm~HDFS1-113}$ & $-2.1\pm1.9$ & $11.2\pm0.9$ & $21.1\pm0.6$ & $38.7\pm1.2$ & $186.3\pm11.5$ & $218.6\pm22.0$ & $354.2\pm18.5$ & $342.4\pm20.0$ \\ 
${\rm~HDFS1-117}$ & $5.8\pm1.9$ & $5.8\pm0.9$ & $6.2\pm0.6$ & $12.9\pm1.2$ & $57.5\pm11.4$ & $148.6\pm21.9$ & $183.8\pm18.5$ & $170.4\pm17.1$ \\ 
${\rm~HDFS1-115}$ & $22.6\pm1.9$ & $49.8\pm0.9$ & $114.4\pm0.6$ & $228.3\pm1.2$ & $373.3\pm11.5$ & $470.6\pm22.0$ & $568.7\pm18.6$ & $564.2\pm22.5$ \\ 
${\rm~HDFS1-127}$ & $25.7\pm1.9$ & $38.5\pm0.9$ & $55.5\pm0.6$ & $113.3\pm1.2$ & $147.7\pm11.5$ & $129.4\pm22.0$ & $192.0\pm18.5$ & $276.3\pm35.2$ \\ 
${\rm~HDFS1-121}$ & $48.4\pm2.0$ & $63.8\pm1.0$ & $117.0\pm0.6$ & $181.6\pm1.2$ & $228.9\pm11.5$ & $204.1\pm22.1$ & $256.1\pm18.6$ & $276.5\pm21.9$ \\ 
${\rm~HDFS1-125}$ & $1.8\pm1.8$ & $7.2\pm0.9$ & $11.8\pm0.6$ & $23.5\pm1.1$ & $99.5\pm10.5$ & $131.5\pm20.1$ & $179.3\pm16.9$ & $178.7\pm23.6$ \\ 
${\rm~HDFS1-131}$ & $5.2\pm1.9$ & $19.0\pm0.9$ & $28.0\pm0.6$ & $45.7\pm1.2$ & $141.8\pm11.5$ & $179.4\pm22.0$ & $244.3\pm18.5$ & $233.5\pm19.7$ \\ 
${\rm~HDFS1-139}$ & $5.1\pm1.9$ & $38.5\pm1.0$ & $54.9\pm0.6$ & $77.7\pm1.2$ & $226.1\pm11.5$ & $294.4\pm22.0$ & $395.2\pm18.5$ & $393.3\pm26.0$ \\ 
${\rm~HDFS1-141}$ & $72.3\pm1.9$ & $103.2\pm1.0$ & $183.8\pm0.7$ & $293.6\pm1.2$ & $381.5\pm11.5$ & $353.1\pm22.0$ & $440.7\pm18.5$ & $577.6\pm29.6$ \\ 
${\rm~HDFS1-148}$ & $-4.1\pm1.9$ & $2.0\pm0.9$ & $3.8\pm0.6$ & $9.6\pm1.2$ & $79.8\pm11.5$ & $191.7\pm22.0$ & $244.4\pm18.6$ & $226.6\pm17.2$ \\ 
${\rm~HDFS1-152}$ & $-1.9\pm1.9$ & $11.1\pm0.9$ & $31.4\pm0.6$ & $40.7\pm1.2$ & $70.6\pm11.5$ & $171.0\pm22.0$ & $176.0\pm18.5$ & $219.0\pm26.6$ \\ 
${\rm~HDFS1-160}$ & $2.5\pm1.9$ & $97.1\pm0.9$ & $150.4\pm0.6$ & $176.0\pm1.1$ & $199.7\pm11.1$ & $286.4\pm21.4$ & $365.2\pm18.0$ & $356.9\pm22.1$ \\ 
${\rm~HDFS1-163}$ & $11.3\pm1.9$ & $42.2\pm0.9$ & $55.6\pm0.6$ & $106.4\pm1.2$ & $402.9\pm11.5$ & $601.6\pm22.0$ & $781.1\pm18.5$ & $812.1\pm26.3$ \\ 
${\rm~HDFS1-173}$ & $19.0\pm1.9$ & $29.0\pm0.9$ & $32.9\pm0.6$ & $52.9\pm1.2$ & $106.4\pm11.5$ & $88.8\pm22.0$ & $148.9\pm18.5$ & $182.9\pm33.7$ \\ 
${\rm~HDFS1-182}$ & $2.8\pm2.0$ & $0.3\pm1.0$ & $3.5\pm0.7$ & $7.0\pm1.2$ & $85.2\pm11.5$ & $211.3\pm22.0$ & $279.8\pm18.5$ & $261.7\pm18.1$ \\ 
${\rm~HDFS1-186}$ & $89.2\pm1.9$ & $234.9\pm0.9$ & $577.6\pm0.6$ & $963.1\pm1.2$ & $1605.1\pm11.5$ & $1933.2\pm22.0$ & $2194.5\pm18.5$ & $2992.2\pm37.2$ \\ 
${\rm~HDFS1-194}$ & $4.5\pm2.0$ & $12.6\pm1.0$ & $35.3\pm0.6$ & $48.1\pm1.2$ & $109.0\pm11.5$ & $156.7\pm22.0$ & $289.2\pm18.5$ & $295.6\pm27.3$ \\ 
${\rm~HDFS1-187}$ & $32.5\pm1.9$ & $71.2\pm0.9$ & $136.3\pm0.6$ & $317.6\pm1.2$ & $858.0\pm11.5$ & $1201.0\pm22.0$ & $1794.9\pm18.5$ & $2208.2\pm32.7$ \\ 
${\rm~HDFS1-188}$ & $19.8\pm1.6$ & $31.8\pm0.8$ & $66.5\pm0.5$ & $130.6\pm1.0$ & $189.0\pm9.9$ & $245.9\pm19.1$ & $248.3\pm16.0$ & $237.8\pm17.0$ \\ 
${\rm~HDFS1-207}$ & $25.6\pm2.0$ & $116.9\pm1.0$ & $478.8\pm0.6$ & $1439.3\pm1.2$ & $3169.8\pm11.5$ & $4177.3\pm22.0$ & $4909.9\pm18.6$ & $5594.4\pm34.5$ \\ 
${\rm~HDFS1-232}$ & $94.7\pm2.1$ & $149.3\pm1.0$ & $308.3\pm0.7$ & $498.4\pm1.2$ & $630.5\pm11.5$ & $728.9\pm22.0$ & $715.6\pm18.5$ & $735.2\pm24.3$ \\ 
${\rm~HDFS1-236}$ & $42.8\pm2.4$ & $59.9\pm1.2$ & $112.0\pm0.7$ & $179.1\pm1.4$ & $244.1\pm11.5$ & $244.9\pm22.0$ & $281.0\pm18.5$ & $301.7\pm33.7$ \\ 
${\rm~HDFS1-237}$ & $40.2\pm2.4$ & $73.1\pm1.1$ & $153.6\pm0.7$ & $298.7\pm1.4$ & $418.8\pm11.5$ & $583.9\pm22.1$ & $576.1\pm18.6$ & $721.1\pm29.2$ \\ 
${\rm~HDFS1-276}$ & $13.1\pm2.3$ & $38.3\pm1.1$ & $46.9\pm0.7$ & $78.6\pm1.5$ & $189.8\pm11.5$ & $247.7\pm22.0$ & $287.5\pm18.5$ & $292.4\pm24.9$ \\ 
${\rm~HDFS1-283}$ & $12.2\pm2.3$ & $30.5\pm1.1$ & $43.1\pm0.7$ & $70.9\pm1.4$ & $158.1\pm11.5$ & $256.2\pm22.0$ & $232.2\pm18.5$ & $219.4\pm19.3$ \\ 
${\rm~HDFS1-286}$ & $17.6\pm3.3$ & $51.4\pm1.6$ & $57.1\pm0.9$ & $84.4\pm1.9$ & $152.3\pm11.5$ & $213.0\pm22.0$ & $218.4\pm18.5$ & $206.2\pm17.7$ \\ 
${\rm~HDFS1-287}$ & $64.9\pm2.5$ & $94.3\pm1.3$ & $132.8\pm0.8$ & $271.7\pm1.7$ & $491.0\pm10.8$ & $492.8\pm20.8$ & $592.0\pm17.5$ & $622.7\pm23.8$ \\ 
${\rm~HDFS1-302}$ & $12.9\pm2.4$ & $34.0\pm1.1$ & $118.3\pm0.7$ & $307.7\pm1.6$ & $611.2\pm11.5$ & $797.7\pm22.0$ & $1006.4\pm18.5$ & $967.5\pm22.5$ \\ 
${\rm~HDFS1-289}$ & $233.6\pm2.6$ & $498.0\pm1.3$ & $989.8\pm0.8$ & $1917.0\pm1.8$ & $3402.5\pm11.5$ & $4810.0\pm22.0$ & $5790.5\pm18.5$ & $7104.0\pm37.7$ \\ 
${\rm~HDFS1-291}$ & $-0.6\pm2.2$ & $5.4\pm1.0$ & $14.2\pm0.7$ & $72.1\pm1.3$ & $358.2\pm11.5$ & $547.6\pm22.1$ & $733.7\pm18.6$ & $700.0\pm19.6$ \\ 
${\rm~HDFS1-299}$ & $77.2\pm2.1$ & $156.7\pm1.0$ & $346.1\pm0.7$ & $688.6\pm1.2$ & $1139.6\pm11.5$ & $1435.2\pm22.0$ & $1619.5\pm18.5$ & $2788.3\pm44.4$ \\ 
${\rm~HDFS1-306}$ & $2.0\pm2.1$ & $19.8\pm1.0$ & $24.6\pm0.7$ & $40.3\pm1.3$ & $103.1\pm11.5$ & $95.0\pm22.0$ & $181.2\pm18.5$ & $177.3\pm18.0$ \\ 
${\rm~HDFS1-313}$ & $39.5\pm2.1$ & $59.8\pm1.0$ & $117.5\pm0.7$ & $201.4\pm1.2$ & $249.6\pm11.5$ & $328.6\pm22.0$ & $333.4\pm18.5$ & $351.2\pm25.6$ \\ 
${\rm~HDFS1-317}$ & $-3.4\pm2.1$ & $7.8\pm1.1$ & $11.4\pm0.7$ & $31.3\pm1.4$ & $23.4\pm11.3$ & $81.7\pm21.7$ & $109.4\pm18.3$ & $159.1\pm26.4$ \\ 
${\rm~HDFS1-318}$ & $13.2\pm2.1$ & $141.6\pm1.0$ & $546.9\pm0.7$ & $1140.3\pm1.2$ & $2001.0\pm11.5$ & $2422.8\pm22.0$ & $2745.0\pm18.6$ & $3336.3\pm34.5$ \\ 
${\rm~HDFS1-335}$ & $5.6\pm2.3$ & $42.6\pm1.1$ & $58.7\pm0.7$ & $71.1\pm1.4$ & $91.6\pm11.5$ & $213.1\pm22.1$ & $233.1\pm18.6$ & $250.0\pm24.8$ \\ 
${\rm~HDFS1-326}$ & $56.3\pm2.1$ & $69.0\pm1.0$ & $111.8\pm0.7$ & $193.5\pm1.2$ & $175.9\pm11.5$ & $277.8\pm22.0$ & $191.7\pm18.5$ & $172.5\pm17.4$ \\ 
${\rm~HDFS1-332}$ & $61.1\pm2.0$ & $170.2\pm1.0$ & $415.1\pm0.7$ & $725.6\pm1.2$ & $1140.3\pm11.5$ & $1462.2\pm22.0$ & $1681.7\pm18.5$ & $1727.9\pm26.5$ \\ 
${\rm~HDFS1-334}$ & $10.1\pm2.1$ & $38.1\pm1.0$ & $86.6\pm0.7$ & $231.1\pm1.2$ & $873.6\pm11.5$ & $1478.9\pm22.0$ & $2203.7\pm18.5$ & $2577.4\pm33.0$ \\ 
${\rm~HDFS1-340}$ & $39.9\pm3.8$ & $67.8\pm1.4$ & $131.6\pm0.9$ & $228.2\pm1.8$ & $315.4\pm11.5$ & $350.1\pm22.0$ & $384.4\pm18.6$ & $433.6\pm24.5$ \\ 
${\rm~HDFS1-342}$ & $50.0\pm1.9$ & $69.0\pm1.0$ & $101.2\pm0.6$ & $193.5\pm1.2$ & $254.3\pm11.5$ & $259.2\pm22.0$ & $307.0\pm18.6$ & $297.6\pm19.7$ \\ 
${\rm~HDFS1-346}$ & $28.6\pm2.1$ & $47.1\pm1.0$ & $94.0\pm0.7$ & $143.0\pm1.2$ & $161.6\pm11.5$ & $187.3\pm22.0$ & $214.6\pm18.5$ & $280.7\pm32.4$ \\ 
${\rm~HDFS1-347}$ & $5.2\pm2.1$ & $19.7\pm1.0$ & $41.1\pm0.7$ & $52.9\pm1.2$ & $60.7\pm11.5$ & $162.8\pm22.0$ & $179.3\pm18.5$ & $214.5\pm27.1$ \\ 
${\rm~HDFS1-345}$ & $41.1\pm1.9$ & $207.3\pm1.0$ & $821.0\pm0.7$ & $2147.5\pm1.2$ & $5113.4\pm11.5$ & $8029.0\pm22.1$ & $10027.0\pm18.6$ & $12210.0\pm38.4$ \\ 
${\rm~HDFS1-350}$ & $-0.2\pm2.0$ & $3.9\pm1.0$ & $9.1\pm0.7$ & $19.0\pm1.2$ & $90.0\pm14.7$ & $169.1\pm27.8$ & $373.7\pm23.7$ & $393.7\pm28.1$ \\ 
${\rm~HDFS1-355}$ & $3.9\pm2.1$ & $4.6\pm1.0$ & $5.7\pm0.7$ & $12.2\pm1.2$ & $12.2\pm11.5$ & $106.0\pm22.0$ & $169.3\pm18.5$ & $165.1\pm17.4$ \\ 
${\rm~HDFS1-354}$ & $3.4\pm1.6$ & $9.6\pm0.8$ & $12.0\pm0.5$ & $22.1\pm1.0$ & $77.4\pm9.4$ & $203.5\pm18.0$ & $155.8\pm15.2$ & $151.6\pm21.3$ \\ 
${\rm~HDFS1-364}$ & $81.1\pm1.9$ & $117.4\pm0.9$ & $164.0\pm0.6$ & $274.9\pm1.2$ & $388.1\pm11.4$ & $452.9\pm21.9$ & $515.8\pm18.5$ & $606.4\pm35.9$ \\ 
${\rm~HDFS1-363}$ & $50.1\pm2.0$ & $115.3\pm1.0$ & $158.0\pm0.7$ & $268.8\pm1.2$ & $580.5\pm11.5$ & $705.6\pm22.0$ & $791.4\pm18.5$ & $781.4\pm21.8$ \\ 
${\rm~HDFS1-360}$ & $27.1\pm1.9$ & $86.4\pm1.0$ & $114.0\pm0.6$ & $198.9\pm1.2$ & $561.7\pm11.4$ & $776.6\pm21.9$ & $817.3\pm18.4$ & $1071.2\pm34.3$ \\ 
${\rm~HDFS1-368}$ & $36.1\pm1.9$ & $60.0\pm0.9$ & $67.7\pm0.6$ & $123.8\pm1.2$ & $195.2\pm11.5$ & $190.1\pm22.0$ & $232.7\pm18.5$ & $250.7\pm23.4$ \\ 
${\rm~HDFS1-372}$ & $32.4\pm1.8$ & $75.8\pm0.9$ & $172.0\pm0.6$ & $336.5\pm1.1$ & $483.2\pm10.4$ & $593.5\pm20.0$ & $700.0\pm16.9$ & $754.1\pm23.2$ \\ 
${\rm~HDFS1-373}$ & $34.1\pm1.8$ & $70.7\pm0.9$ & $112.0\pm0.6$ & $202.1\pm1.1$ & $388.1\pm10.3$ & $445.5\pm19.7$ & $595.3\pm16.6$ & $607.9\pm21.9$ \\ 
${\rm~HDFS1-378}$ & $9.2\pm2.0$ & $72.6\pm1.0$ & $98.4\pm0.7$ & $130.4\pm1.2$ & $311.1\pm11.5$ & $422.2\pm22.1$ & $499.1\pm18.6$ & $648.2\pm40.3$ \\ 
${\rm~HDFS1-379}$ & $38.0\pm2.1$ & $67.8\pm1.0$ & $124.7\pm0.7$ & $345.9\pm1.2$ & $1373.1\pm11.5$ & $2079.8\pm22.0$ & $2726.2\pm18.5$ & $2881.6\pm28.2$ \\ 
${\rm~HDFS1-377}$ & $2.1\pm2.1$ & $15.7\pm1.0$ & $34.2\pm0.7$ & $130.2\pm1.3$ & $700.4\pm11.5$ & $1090.0\pm22.0$ & $1551.8\pm18.5$ & $1590.3\pm24.9$ \\ 
${\rm~HDFS1-380}$ & $-3.1\pm1.9$ & $8.8\pm0.9$ & $23.7\pm0.6$ & $101.9\pm1.2$ & $291.4\pm11.5$ & $408.5\pm22.0$ & $518.0\pm18.6$ & $478.4\pm20.0$ \\ 
${\rm~HDFS1-381}$ & $18.4\pm1.9$ & $37.3\pm0.9$ & $45.5\pm0.6$ & $83.8\pm1.2$ & $121.8\pm11.5$ & $184.8\pm22.0$ & $219.2\pm18.5$ & $213.9\pm21.5$ \\ 
${\rm~HDFS1-382}$ & $-2.7\pm1.9$ & $24.9\pm0.9$ & $31.2\pm0.6$ & $45.8\pm1.2$ & $70.6\pm11.4$ & $121.7\pm21.9$ & $191.8\pm18.4$ & $206.7\pm24.9$ \\ 
${\rm~HDFS1-386}$ & $13.7\pm1.9$ & $102.0\pm0.9$ & $139.6\pm0.6$ & $191.4\pm1.2$ & $462.9\pm11.5$ & $575.7\pm22.0$ & $620.5\pm18.5$ & $665.6\pm26.6$ \\ 
${\rm~HDFS1-383}$ & $65.3\pm1.9$ & $119.3\pm0.9$ & $266.5\pm0.6$ & $412.6\pm1.2$ & $543.2\pm11.5$ & $642.0\pm22.0$ & $656.8\pm18.5$ & $1037.5\pm44.3$ \\ 
${\rm~HDFS1-424}$ & $0.0\pm1.9$ & $2.3\pm0.9$ & $7.6\pm0.6$ & $35.7\pm1.2$ & $31.3\pm11.5$ & $48.8\pm22.0$ & $152.9\pm18.5$ & $177.3\pm35.0$ \\ 
${\rm~HDFS1-393}$ & $5.0\pm1.9$ & $33.0\pm0.9$ & $47.7\pm0.6$ & $75.6\pm1.2$ & $284.3\pm11.5$ & $352.9\pm22.0$ & $544.6\pm18.5$ & $664.9\pm30.4$ \\ 
${\rm~HDFS1-394}$ & $4.5\pm1.9$ & $25.6\pm1.0$ & $63.6\pm0.6$ & $109.9\pm1.2$ & $161.7\pm13.0$ & $198.9\pm25.0$ & $188.0\pm21.1$ & $308.1\pm41.0$ \\ 
${\rm~HDFS1-395}$ & $73.1\pm2.1$ & $291.9\pm1.0$ & $603.1\pm0.7$ & $930.9\pm1.2$ & $1321.3\pm11.5$ & $1516.6\pm22.1$ & $1554.7\pm18.6$ & $1947.7\pm33.0$ \\ 
${\rm~HDFS1-397}$ & $14.4\pm2.1$ & $23.4\pm1.0$ & $38.7\pm0.7$ & $94.8\pm1.2$ & $323.4\pm11.5$ & $526.9\pm22.1$ & $778.5\pm18.6$ & $952.8\pm34.7$ \\ 
${\rm~HDFS1-399}$ & $57.9\pm2.0$ & $89.7\pm1.0$ & $180.2\pm0.7$ & $315.6\pm1.2$ & $410.7\pm11.5$ & $532.8\pm22.0$ & $568.3\pm18.5$ & $546.5\pm23.4$ \\ 
${\rm~HDFS1-404}$ & $27.3\pm2.9$ & $43.4\pm1.2$ & $70.1\pm0.8$ & $112.3\pm1.5$ & $151.1\pm11.5$ & $153.3\pm22.0$ & $186.8\pm18.6$ & $261.2\pm33.6$ \\ 
${\rm~HDFS1-405}$ & $5.1\pm1.9$ & $3.3\pm0.9$ & $9.2\pm0.6$ & $47.5\pm1.2$ & $258.1\pm11.5$ & $341.2\pm22.0$ & $541.7\pm18.5$ & $565.7\pm25.5$ \\ 
${\rm~HDFS1-398}$ & $1.1\pm1.9$ & $1.8\pm0.9$ & $5.0\pm0.6$ & $24.1\pm1.1$ & $84.8\pm10.7$ & $141.2\pm20.4$ & $231.5\pm17.2$ & $335.1\pm31.0$ \\ 
${\rm~HDFS1-406}$ & $118.8\pm2.1$ & $343.0\pm1.0$ & $1174.0\pm0.7$ & $3242.7\pm1.2$ & $7359.3\pm11.5$ & $10974.2\pm22.0$ & $13863.9\pm18.5$ & $21541.4\pm49.5$ \\ 
${\rm~HDFS1-411}$ & $11.1\pm2.2$ & $6.6\pm1.1$ & $16.2\pm0.7$ & $41.3\pm1.3$ & $134.5\pm11.5$ & $235.4\pm22.0$ & $361.3\pm18.5$ & $370.4\pm28.9$ \\ 
${\rm~HDFS1-427}$ & $10.5\pm1.9$ & $16.1\pm0.9$ & $22.5\pm0.6$ & $42.8\pm1.2$ & $148.6\pm11.5$ & $186.9\pm22.0$ & $264.1\pm18.5$ & $261.3\pm20.2$ \\ 
${\rm~HDFS1-414}$ & $26.6\pm2.0$ & $89.1\pm1.0$ & $236.7\pm0.7$ & $526.1\pm1.2$ & $1248.8\pm11.5$ & $2030.9\pm22.1$ & $2714.7\pm18.6$ & $2770.6\pm26.2$ \\ 
${\rm~HDFS1-410}$ & $85.3\pm2.0$ & $144.6\pm1.0$ & $285.7\pm0.7$ & $491.4\pm1.2$ & $671.5\pm11.5$ & $782.2\pm22.0$ & $882.8\pm18.5$ & $1481.1\pm41.1$ \\ 
${\rm~HDFS1-415}$ & $61.0\pm1.9$ & $116.1\pm0.9$ & $252.7\pm0.7$ & $415.9\pm1.2$ & $552.8\pm11.5$ & $673.8\pm22.0$ & $637.9\pm18.5$ & $839.5\pm33.0$ \\ 
${\rm~HDFS1-421}$ & $121.1\pm1.9$ & $428.1\pm0.9$ & $1243.9\pm0.6$ & $2490.1\pm1.2$ & $4647.2\pm11.6$ & $6378.8\pm22.3$ & $7614.6\pm18.7$ & $9886.4\pm40.3$ \\ 
${\rm~HDFS1-426}$ & $13.0\pm2.1$ & $15.4\pm1.0$ & $33.9\pm0.7$ & $85.1\pm1.2$ & $259.9\pm11.5$ & $465.1\pm22.1$ & $659.7\pm18.6$ & $639.4\pm21.6$ \\ 
${\rm~HDFS1-434}$ & $15.6\pm2.0$ & $15.9\pm1.0$ & $41.1\pm0.7$ & $84.0\pm1.2$ & $119.2\pm11.5$ & $170.4\pm22.0$ & $230.1\pm18.5$ & $250.1\pm20.6$ \\ 
${\rm~HDFS1-435}$ & $1.6\pm2.0$ & $13.8\pm1.0$ & $37.6\pm0.7$ & $89.5\pm1.2$ & $139.1\pm11.5$ & $193.4\pm22.0$ & $227.9\pm18.6$ & $244.0\pm31.9$ \\ 
${\rm~HDFS1-437}$ & $49.1\pm2.0$ & $69.2\pm1.0$ & $81.6\pm0.7$ & $127.7\pm1.2$ & $201.1\pm11.5$ & $226.5\pm22.1$ & $285.0\pm18.6$ & $288.0\pm23.0$ \\ 
${\rm~HDFS1-439}$ & $93.5\pm1.9$ & $140.1\pm0.9$ & $212.6\pm0.6$ & $382.9\pm1.2$ & $496.9\pm11.5$ & $610.1\pm22.0$ & $637.5\pm18.5$ & $851.0\pm36.3$ \\ 
${\rm~HDFS1-440}$ & $6.2\pm1.9$ & $0.1\pm0.9$ & $6.7\pm0.6$ & $33.9\pm1.2$ & $284.0\pm11.5$ & $569.8\pm22.0$ & $781.8\pm18.5$ & $785.1\pm24.6$ \\ 
${\rm~HDFS1-448}$ & $31.5\pm2.3$ & $62.4\pm1.1$ & $71.2\pm0.8$ & $100.2\pm1.4$ & $198.3\pm11.8$ & $264.2\pm22.5$ & $258.6\pm19.0$ & $314.4\pm28.7$ \\ 
${\rm~HDFS1-450}$ & $20.0\pm1.7$ & $27.1\pm0.9$ & $65.5\pm0.6$ & $102.8\pm1.1$ & $88.8\pm10.7$ & $78.6\pm20.6$ & $125.8\pm17.3$ & $167.4\pm29.7$ \\ 
${\rm~HDFS1-463}$ & $-0.5\pm1.9$ & $15.4\pm0.9$ & $20.5\pm0.6$ & $26.9\pm1.2$ & $69.1\pm11.5$ & $87.9\pm22.0$ & $216.9\pm18.6$ & $234.5\pm21.0$ \\ 
${\rm~HDFS1-484}$ & $6.0\pm2.1$ & $64.3\pm1.0$ & $353.4\pm0.7$ & $1138.1\pm1.3$ & $2428.7\pm11.5$ & $3465.8\pm22.0$ & $4114.4\pm18.5$ & $4158.8\pm26.3$ \\ 
${\rm~HDFS1-472}$ & $12.2\pm2.1$ & $36.7\pm1.0$ & $74.9\pm0.7$ & $168.4\pm1.2$ & $253.0\pm11.5$ & $339.9\pm22.0$ & $350.5\pm18.5$ & $367.6\pm21.9$ \\ 
${\rm~HDFS1-476}$ & $27.0\pm1.8$ & $49.8\pm0.9$ & $59.1\pm0.6$ & $97.2\pm1.2$ & $186.9\pm11.5$ & $193.6\pm22.0$ & $210.5\pm18.5$ & $208.1\pm27.9$ \\ 
${\rm~HDFS1-480}$ & $-2.5\pm2.1$ & $2.3\pm1.0$ & $0.3\pm0.7$ & $1.7\pm1.3$ & $16.8\pm11.5$ & $70.4\pm22.0$ & $174.6\pm18.5$ & $171.6\pm20.0$ \\ 
${\rm~HDFS1-479}$ & $5.7\pm1.9$ & $12.1\pm0.9$ & $26.3\pm0.7$ & $47.0\pm1.2$ & $156.2\pm11.5$ & $217.7\pm22.0$ & $378.9\pm18.5$ & $388.5\pm21.4$ \\ 
${\rm~HDFS1-483}$ & $4.4\pm1.8$ & $34.1\pm0.9$ & $43.9\pm0.6$ & $58.8\pm1.2$ & $155.8\pm11.5$ & $185.6\pm22.1$ & $266.2\pm18.6$ & $328.1\pm28.8$ \\ 
${\rm~HDFS1-487}$ & $3.0\pm2.1$ & $1.8\pm1.0$ & $4.4\pm0.7$ & $16.4\pm1.2$ & $55.2\pm11.5$ & $187.5\pm22.0$ & $213.5\pm18.5$ & $196.5\pm17.8$ \\ 
${\rm~HDFS1-488}$ & $42.3\pm2.1$ & $92.3\pm1.0$ & $235.8\pm0.7$ & $493.2\pm1.2$ & $873.6\pm11.5$ & $1245.4\pm22.0$ & $1533.3\pm18.5$ & $1947.3\pm32.3$ \\ 
${\rm~HDFS1-492}$ & $19.7\pm2.1$ & $31.9\pm1.0$ & $44.5\pm0.7$ & $57.7\pm1.3$ & $105.6\pm11.5$ & $100.6\pm22.0$ & $147.5\pm18.5$ & $164.7\pm24.0$ \\ 
${\rm~HDFS1-489}$ & $46.1\pm2.2$ & $96.0\pm1.0$ & $236.3\pm0.7$ & $513.2\pm1.3$ & $934.6\pm11.5$ & $1322.8\pm22.0$ & $1568.4\pm18.6$ & $1647.6\pm26.6$ \\ 
${\rm~HDFS1-478}$ & $47.4\pm2.2$ & $77.4\pm1.0$ & $97.1\pm0.7$ & $176.7\pm1.3$ & $553.1\pm11.5$ & $682.6\pm22.0$ & $907.6\pm18.5$ & $1157.7\pm37.3$ \\ 
${\rm~HDFS1-505}$ & $2.3\pm1.9$ & $35.3\pm1.0$ & $50.0\pm0.6$ & $93.3\pm1.2$ & $265.1\pm11.5$ & $438.4\pm22.0$ & $423.6\pm18.5$ & $425.5\pm20.3$ \\ 
${\rm~HDFS1-511}$ & $28.5\pm2.3$ & $63.3\pm1.1$ & $97.3\pm0.7$ & $190.9\pm1.3$ & $451.0\pm11.5$ & $669.7\pm22.1$ & $889.9\pm18.6$ & $962.4\pm26.5$ \\ 
${\rm~HDFS1-516}$ & $48.9\pm3.4$ & $68.8\pm1.4$ & $126.7\pm0.9$ & $188.2\pm1.7$ & $205.6\pm11.5$ & $268.6\pm22.0$ & $229.7\pm18.5$ & $381.8\pm37.0$ \\ 
${\rm~HDFS1-542}$ & $3.7\pm2.4$ & $28.3\pm1.1$ & $81.2\pm0.8$ & $139.0\pm1.3$ & $216.4\pm11.5$ & $195.7\pm22.0$ & $291.6\pm18.5$ & $293.5\pm23.1$ \\ 
${\rm~HDFS1-521}$ & $80.4\pm2.8$ & $177.6\pm1.3$ & $427.4\pm0.9$ & $875.4\pm1.5$ & $1620.2\pm11.6$ & $2313.2\pm22.3$ & $2721.3\pm18.8$ & $3263.4\pm35.4$ \\ 
${\rm~HDFS1-522}$ & $13.2\pm2.0$ & $29.7\pm1.0$ & $58.3\pm0.7$ & $109.2\pm1.2$ & $145.8\pm15.6$ & $111.6\pm29.3$ & $232.5\pm25.1$ & $222.4\pm26.0$ \\ 
${\rm~HDFS1-530}$ & $26.2\pm3.5$ & $52.3\pm1.6$ & $73.6\pm1.0$ & $158.6\pm1.8$ & $357.5\pm11.5$ & $395.9\pm22.0$ & $520.6\pm18.5$ & $531.7\pm24.6$ \\ 
${\rm~HDFS1-536}$ & $25.1\pm1.9$ & $43.8\pm1.0$ & $66.0\pm0.7$ & $144.9\pm1.2$ & $203.9\pm11.5$ & $257.9\pm22.0$ & $258.4\pm18.5$ & $294.0\pm32.1$ \\ 
${\rm~HDFS1-527}$ & $31.0\pm2.0$ & $59.3\pm1.0$ & $104.5\pm0.7$ & $299.5\pm1.2$ & $1278.7\pm11.5$ & $2125.7\pm22.0$ & $2883.0\pm18.6$ & $3437.7\pm35.1$ \\ 
${\rm~HDFS1-538}$ & $97.8\pm2.4$ & $162.6\pm1.1$ & $294.2\pm0.8$ & $504.3\pm1.4$ & $706.7\pm11.5$ & $962.7\pm22.1$ & $1025.3\pm18.6$ & $1059.7\pm25.0$ \\ 
${\rm~HDFS1-548}$ & $12.9\pm2.3$ & $8.5\pm1.1$ & $41.6\pm0.8$ & $174.2\pm1.3$ & $508.4\pm11.5$ & $787.7\pm22.1$ & $835.5\pm18.6$ & $922.8\pm30.2$ \\ 
${\rm~HDFS1-555}$ & $-5.9\pm3.0$ & $8.5\pm1.5$ & $22.7\pm1.1$ & $95.6\pm1.6$ & $427.0\pm12.7$ & $743.0\pm24.5$ & $1202.9\pm20.5$ & $1204.0\pm26.4$ \\ 
\label{tbl:table1.in}
\enddata
\tablenotetext{a} {Fluxes measured over a 2\farcs0 diameter aperture.}
\tablenotetext{b} {The ``AUTO'' flux from SExtractor with a minimum 2\farcs0 diameter aperture.}
\tablecomments{All fluxes in units of $10^{-31}~{\rm ergs~s^{-1}Hz^{-1}cm^{-2}}$.
}
\end{deluxetable}

\clearpage

\begin{deluxetable}{llll}
\tablewidth{0pt}
\tablecaption{NIR~Template~Extension~Parameters}
\tablehead{\colhead{Template} & \colhead{Age} & \colhead{IMF} & \colhead{SFR}\\
\colhead{ } & \colhead{Gyr} & \colhead{ } & \colhead{ }}
\startdata
E/S0 & $12.7$ & Scalo & $\tau = 1$ Gyr \\
Sbc & $12.7$ & Scalo & $\tau = 8$ Gyr \\
Scd & $12.7$ & Salpeter & Constant \\
Irr & $0.1$ & Salpeter & Constant \\
SB1 & $0.1$ & Salpeter & Constant \\
SB2 & $0.1$ & Salpeter & Constant \\
\enddata
\end{deluxetable}

\clearpage

\begin{deluxetable}{lllllll}
\tablewidth{0pt}
\tablecaption{Photometric~Redshift~Catalog}
\tablehead{\colhead{ID} & \colhead{RA (22h)} & \colhead{DEC ($-60^{\circ}$)} & \colhead{$z_{phot}$} & \colhead{${\rm L_U^{rest}}$} & \colhead{${\rm L_B^{rest}}$} & \colhead{${\rm L_V^{rest}}$
}\\
\colhead{ } & \colhead{J2000} & \colhead{J2000} & \colhead{ } & \colhead{$10^{10}~{\rm L_{\odot}}$} & \colhead{$10^{10}~{\rm L_{\odot}}$} & \colhead{$10^{10}~{\rm L_{\odot}}$
}}
\startdata
${\rm~HDFS1-30}$ & $32:52.26$ & $31:52.7$ & $1.36^{0.17}_{0.17}$ & $3.74^{1.43}_{1.40}$ & $2.53^{1.00}_{0.94}$ & $2.36^{0.96}_{0.86}$ \\ 
${\rm~HDFS1-33}$ & $32:52.69$ & $31:53.0$ & $0.92^{0.13}_{0.14}$ & $0.59^{0.27}_{0.25}$ & $0.34^{0.18}_{0.13}$ & $0.29^{0.17}_{0.11}$ \\ 
${\rm~HDFS1-31}$ & $32:52.04$ & $31:54.1$ & $0.62^{0.15}_{0.13}\tablenotemark{a}$ & $0.12^{0.09}_{0.06}$ & $0.07^{0.05}_{0.04}$ & $0.07^{0.05}_{0.03}$ \\ 
${\rm~HDFS1-36}$ & $32:48.84$ & $31:54.1$ & $3.32^{0.30}_{0.31}$ & $16.53^{4.64}_{0.98}$ & $9.21^{2.50}_{0.38}$ & $8.13^{3.08}_{0.49}$ \\ 
${\rm~HDFS1-37}$ & $32:53.38$ & $31:54.5$ & $3.00^{0.63}_{0.37}\tablenotemark{a}$ & $5.06^{4.47}_{1.92}$ & $5.25^{4.64}_{1.99}$ & $6.62^{5.88}_{2.50}$ \\ 
${\rm~HDFS1-45}\tablenotemark{b}$ & $32:56.18$ & $31:56.6$ & $5.34^{0.44}_{0.45}$ & $59.89^{11.84}_{14.96}$ & $32.92^{0.11}_{6.75}$ & $27.59^{0.08}_{6.67}$ \\ 
${\rm~HDFS1-50}$ & $32:49.45$ & $31:58.1$ & $1.22^{0.28}_{0.16}$ & $0.28^{0.40}_{0.12}$ & $0.32^{0.33}_{0.13}$ & $0.43^{0.35}_{0.17}$ \\ 
${\rm~HDFS1-52}$ & $32:54.06$ & $31:58.1$ & $1.22^{0.16}_{0.18}$ & $1.57^{0.49}_{0.51}$ & $0.89^{0.24}_{0.33}$ & $0.69^{0.19}_{0.25}$ \\ 
${\rm~HDFS1-54}$ & $32:52.98$ & $31:58.4$ & $1.08^{0.15}_{0.28}$ & $0.53^{0.25}_{0.29}$ & $0.32^{0.15}_{0.20}$ & $0.27^{0.15}_{0.16}$ \\ 
${\rm~HDFS1-62}$ & $32:50.35$ & $32:01.0$ & $1.00^{0.14}_{0.17}$ & $0.63^{0.29}_{0.26}$ & $0.40^{0.21}_{0.18}$ & $0.35^{0.19}_{0.16}$ \\ 
${\rm~HDFS1-58}$ & $32:53.38$ & $32:01.3$ & $1.02^{0.14}_{0.14}$ & $0.52^{0.39}_{0.26}$ & $0.60^{0.30}_{0.27}$ & $0.79^{0.31}_{0.33}$ \\ 
${\rm~HDFS1-63}$ & $32:50.28$ & $32:03.5$ & $0.44^{0.10}_{0.10}$ & $0.40^{0.26}_{0.21}$ & $0.26^{0.16}_{0.13}$ & $0.25^{0.13}_{0.12}$ \\ 
${\rm~HDFS1-69}$ & $32:48.80$ & $32:03.5$ & $0.84^{0.24}_{0.13}$ & $0.43^{0.48}_{0.18}$ & $0.25^{0.32}_{0.10}$ & $0.24^{0.29}_{0.08}$ \\ 
${\rm~HDFS1-74}$ & $32:53.70$ & $32:06.0$ & $0.96^{0.14}_{0.14}$ & $1.01^{0.85}_{0.48}$ & $0.90^{0.58}_{0.40}$ & $1.05^{0.50}_{0.42}$ \\ 
${\rm~HDFS1-79}$ & $32:49.06$ & $32:06.0$ & $2.22^{0.23}_{0.24}$ & $10.13^{2.32}_{2.31}$ & $5.54^{0.99}_{1.20}$ & $4.26^{0.81}_{0.94}$ \\ 
${\rm~HDFS1-80}$ & $32:51.86$ & $32:06.0$ & $3.24^{0.30}_{0.30}$ & $8.00^{1.48}_{0.67}$ & $4.07^{0.28}_{0.23}$ & $3.21^{0.26}_{0.14}$ \\ 
${\rm~HDFS1-83}$ & $32:52.73$ & $32:07.1$ & $0.46^{0.10}_{0.10}$ & $0.62^{0.40}_{0.30}$ & $0.39^{0.20}_{0.19}$ & $0.35^{0.17}_{0.15}$ \\ 
${\rm~HDFS1-86}$ & $32:46.68$ & $32:07.1$ & $0.16^{0.08}_{0.08}$ & $0.02^{0.03}_{0.01}$ & $0.01^{0.02}_{0.01}$ & $0.01^{0.02}_{0.01}$ \\ 
${\rm~HDFS1-87}$ & $32:54.82$ & $32:08.2$ & $1.60^{1.93}_{0.21}\tablenotemark{a}$ & $0.87^{6.98}_{0.30}$ & $0.63^{5.03}_{0.23}$ & $0.62^{4.87}_{0.23}$ \\ 
${\rm~HDFS1-92}$ & $32:56.26$ & $32:09.6$ & $1.38^{0.18}_{0.18}$ & $0.99^{0.40}_{0.40}$ & $0.68^{0.28}_{0.25}$ & $0.63^{0.27}_{0.21}$ \\ 
${\rm~HDFS1-98}$ & $32:55.72$ & $32:11.4$ & $0.56^{0.11}_{0.11}$ & $0.45^{0.46}_{0.23}$ & $0.58^{0.40}_{0.29}$ & $0.80^{0.46}_{0.40}$ \\ 
${\rm~HDFS1-105}$ & $32:49.24$ & $32:11.8$ & $2.14^{0.22}_{0.27}$ & $1.70^{0.67}_{0.73}$ & $1.91^{0.67}_{0.73}$ & $2.50^{0.82}_{0.89}$ \\ 
${\rm~HDFS1-107}$ & $32:51.65$ & $32:12.5$ & $1.00^{0.14}_{0.15}$ & $0.79^{0.39}_{0.25}$ & $0.46^{0.19}_{0.19}$ & $0.37^{0.19}_{0.15}$ \\ 
${\rm~HDFS1-99}$ & $32:55.75$ & $32:13.6$ & $0.72^{0.13}_{0.12}$ & $0.29^{0.17}_{0.13}$ & $0.19^{0.10}_{0.08}$ & $0.16^{0.09}_{0.06}$ \\ 
${\rm~HDFS1-119}$ & $32:52.01$ & $32:15.0$ & $0.84^{0.13}_{0.13}$ & $0.77^{0.35}_{0.31}$ & $0.49^{0.23}_{0.19}$ & $0.43^{0.21}_{0.16}$ \\ 
${\rm~HDFS1-111}$ & $32:54.82$ & $32:15.4$ & $0.52^{0.11}_{0.11}$ & $0.38^{0.41}_{0.21}$ & $0.45^{0.33}_{0.24}$ & $0.61^{0.35}_{0.31}$ \\ 
${\rm~HDFS1-112}$ & $32:54.42$ & $32:15.4$ & $2.14^{0.22}_{0.23}$ & $4.15^{0.97}_{0.85}$ & $2.35^{0.47}_{0.49}$ & $1.79^{0.37}_{0.34}$ \\ 
${\rm~HDFS1-113}$ & $32:52.58$ & $32:15.4$ & $1.50^{0.18}_{0.18}$ & $1.45^{0.53}_{0.52}$ & $1.08^{0.40}_{0.40}$ & $1.08^{0.39}_{0.40}$ \\ 
${\rm~HDFS1-117}$ & $32:52.91$ & $32:15.7$ & $1.54^{0.23}_{0.25}$ & $0.44^{0.25}_{0.18}$ & $0.39^{0.22}_{0.17}$ & $0.48^{0.25}_{0.21}$ \\ 
${\rm~HDFS1-115}$ & $32:48.88$ & $32:16.1$ & $0.54^{0.11}_{0.11}$ & $0.26^{0.17}_{0.13}$ & $0.19^{0.10}_{0.09}$ & $0.18^{0.09}_{0.08}$ \\ 
${\rm~HDFS1-127}$ & $32:53.05$ & $32:17.2$ & $0.78^{0.13}_{0.13}$ & $0.60^{0.27}_{0.25}$ & $0.37^{0.17}_{0.15}$ & $0.31^{0.14}_{0.12}$ \\ 
${\rm~HDFS1-121}$ & $32:55.54$ & $32:17.5$ & $0.48^{0.10}_{0.10}$ & $0.22^{0.12}_{0.10}$ & $0.14^{0.07}_{0.06}$ & $0.11^{0.05}_{0.05}$ \\ 
${\rm~HDFS1-125}$ & $32:48.16$ & $32:18.2$ & $1.40^{0.18}_{0.19}$ & $0.68^{0.29}_{0.31}$ & $0.50^{0.21}_{0.20}$ & $0.49^{0.21}_{0.17}$ \\ 
${\rm~HDFS1-131}$ & $32:52.08$ & $32:18.6$ & $1.38^{0.18}_{0.18}\tablenotemark{a}$ & $1.10^{0.44}_{0.43}$ & $0.75^{0.30}_{0.28}$ & $0.70^{0.29}_{0.24}$ \\ 
${\rm~HDFS1-139}$ & $32:47.80$ & $32:19.7$ & $2.24^{0.25}_{0.23}$ & $7.16^{1.93}_{1.69}$ & $4.52^{1.42}_{0.89}$ & $4.16^{1.22}_{0.83}$ \\ 
${\rm~HDFS1-141}$ & $32:56.08$ & $32:20.4$ & $0.50^{0.11}_{0.11}$ & $0.48^{0.28}_{0.23}$ & $0.29^{0.15}_{0.15}$ & $0.24^{0.12}_{0.11}$ \\ 
${\rm~HDFS1-148}$ & $32:50.50$ & $32:22.6$ & $1.72^{0.22}_{0.23}$ & $0.74^{0.38}_{0.33}$ & $0.73^{0.33}_{0.30}$ & $0.89^{0.38}_{0.33}$ \\ 
${\rm~HDFS1-152}$ & $32:52.01$ & $32:24.4$ & $3.50^{0.33}_{0.35}$ & $11.65^{2.88}_{1.83}$ & $7.12^{2.31}_{1.18}$ & $5.94^{4.15}_{0.73}$ \\ 
${\rm~HDFS1-160}$ & $32:49.16$ & $32:26.2$ & $3.00^{0.28}_{0.28}$ & $22.84^{4.28}_{3.03}$ & $10.85^{1.58}_{0.72}$ & $8.77^{1.18}_{0.43}$ \\ 
${\rm~HDFS1-163}$ & $32:48.44$ & $32:28.7$ & $1.42^{0.17}_{0.17}$ & $3.32^{1.18}_{1.35}$ & $2.39^{0.85}_{0.89}$ & $2.34^{0.84}_{0.76}$ \\ 
${\rm~HDFS1-173}$ & $32:53.52$ & $32:31.9$ & $1.12^{0.18}_{0.15}$ & $0.81^{0.41}_{0.27}$ & $0.45^{0.22}_{0.16}$ & $0.37^{0.21}_{0.12}$ \\ 
${\rm~HDFS1-182}$ & $32:46.79$ & $32:33.7$ & $1.82^{0.21}_{0.24}$ & $0.87^{0.40}_{0.41}$ & $0.93^{0.38}_{0.39}$ & $1.19^{0.44}_{0.45}$ \\ 
${\rm~HDFS1-186}$ & $32:53.66$ & $32:35.9$ & $0.20^{0.09}_{0.08}$ & $0.10^{0.16}_{0.08}$ & $0.08^{0.11}_{0.06}$ & $0.09^{0.12}_{0.06}$ \\ 
${\rm~HDFS1-194}$ & $32:48.37$ & $32:38.0$ & $3.52^{0.32}_{0.33}$ & $13.69^{2.61}_{2.73}$ & $9.03^{3.12}_{1.80}$ & $8.16^{5.88}_{1.55}$ \\ 
${\rm~HDFS1-187}$ & $32:53.34$ & $32:39.1$ & $0.90^{0.13}_{0.13}$ & $2.03^{0.97}_{0.85}$ & $1.45^{0.81}_{0.59}$ & $1.56^{0.75}_{0.58}$ \\ 
${\rm~HDFS1-188}$ & $32:53.12$ & $32:39.1$ & $0.58^{0.11}_{0.11}$ & $0.19^{0.11}_{0.09}$ & $0.13^{0.06}_{0.06}$ & $0.12^{0.06}_{0.05}$ \\ 
${\rm~HDFS1-207}$ & $32:50.89$ & $32:43.1$ & $0.54^{0.11}_{0.11}$ & $1.05^{0.92}_{0.66}$ & $1.07^{0.71}_{0.57}$ & $1.34^{0.75}_{0.64}$ \\ 
${\rm~HDFS1-232}$ & $32:54.06$ & $32:51.7$ & $0.48^{0.10}_{0.10}$ & $0.54^{0.31}_{0.26}$ & $0.34^{0.17}_{0.16}$ & $0.30^{0.14}_{0.13}$ \\ 
${\rm~HDFS1-236}$ & $32:47.65$ & $32:52.4$ & $0.50^{0.10}_{0.11}$ & $0.24^{0.15}_{0.11}$ & $0.15^{0.08}_{0.07}$ & $0.12^{0.06}_{0.05}$ \\ 
${\rm~HDFS1-237}$ & $32:49.24$ & $32:53.5$ & $0.58^{0.11}_{0.11}$ & $0.57^{0.33}_{0.26}$ & $0.38^{0.18}_{0.18}$ & $0.35^{0.17}_{0.15}$ \\ 
${\rm~HDFS1-276}$ & $32:51.18$ & $33:01.4$ & $1.26^{0.16}_{0.16}$ & $1.42^{0.46}_{0.51}$ & $0.91^{0.30}_{0.33}$ & $0.81^{0.27}_{0.28}$ \\ 
${\rm~HDFS1-283}$ & $32:47.04$ & $33:02.9$ & $1.20^{0.15}_{0.17}$ & $1.02^{0.35}_{0.41}$ & $0.65^{0.23}_{0.26}$ & $0.58^{0.21}_{0.22}$ \\ 
${\rm~HDFS1-286}$ & $33:0.04$ & $33:04.0$ & $1.24^{0.16}_{0.16}\tablenotemark{a}$ & $1.33^{0.36}_{0.44}$ & $0.76^{0.23}_{0.24}$ & $0.62^{0.18}_{0.20}$ \\ 
${\rm~HDFS1-287}$ & $32:57.26$ & $33:05.4$ & $0.86^{0.13}_{0.13}$ & $1.43^{0.58}_{0.58}$ & $0.86^{0.40}_{0.33}$ & $0.75^{0.34}_{0.28}$ \\ 
${\rm~HDFS1-302}$ & $32:54.02$ & $33:05.4$ & $0.54^{0.11}_{0.11}$ & $0.23^{0.19}_{0.14}$ & $0.21^{0.13}_{0.11}$ & $0.24^{0.12}_{0.11}$ \\ 
${\rm~HDFS1-289}$ & $32:57.59$ & $33:06.1$ & $0.58^{0.11}_{0.11}$ & $3.57^{2.24}_{1.61}$ & $2.42^{1.15}_{1.16}$ & $2.23^{1.10}_{0.92}$ \\ 
${\rm~HDFS1-291}$ & $32:51.68$ & $33:06.1$ & $0.98^{0.14}_{0.14}$ & $0.35^{0.33}_{0.17}$ & $0.42^{0.27}_{0.19}$ & $0.57^{0.28}_{0.25}$ \\ 
${\rm~HDFS1-299}$ & $32:52.30$ & $33:08.3$ & $0.56^{0.11}_{0.11}$ & $1.55^{0.97}_{0.75}$ & $1.08^{0.56}_{0.52}$ & $1.04^{0.51}_{0.45}$ \\ 
${\rm~HDFS1-306}$ & $32:48.05$ & $33:09.4$ & $1.30^{0.16}_{0.17}\tablenotemark{a}$ & $0.81^{0.27}_{0.31}$ & $0.51^{0.20}_{0.19}$ & $0.46^{0.18}_{0.17}$ \\ 
${\rm~HDFS1-313}$ & $32:49.49$ & $33:11.2$ & $0.52^{0.11}_{0.11}$ & $0.27^{0.17}_{0.13}$ & $0.17^{0.08}_{0.08}$ & $0.15^{0.08}_{0.07}$ \\ 
${\rm~HDFS1-317}$ & $33:2.02$ & $33:12.6$ & $0.78^{0.16}_{0.13}\tablenotemark{a}$ & $0.14^{0.11}_{0.07}$ & $0.10^{0.07}_{0.04}$ & $0.09^{0.07}_{0.04}$ \\ 
${\rm~HDFS1-318}$ & $32:53.92$ & $33:13.3$ & $0.20^{0.08}_{0.08}$ & $0.05^{0.10}_{0.03}$ & $0.06^{0.10}_{0.04}$ & $0.08^{0.11}_{0.06}$ \\ 
${\rm~HDFS1-335}$ & $33:4.00$ & $33:13.7$ & $2.54^{0.26}_{0.26}$ & $7.55^{1.44}_{1.50}$ & $4.22^{1.08}_{0.87}$ & $3.49^{0.97}_{0.75}$ \\ 
${\rm~HDFS1-326}$ & $32:48.55$ & $33:14.0$ & $0.62^{0.13}_{0.12}$ & $0.37^{0.25}_{0.15}$ & $0.22^{0.12}_{0.10}$ & $0.18^{0.10}_{0.07}$ \\ 
${\rm~HDFS1-332}$ & $33:1.94$ & $33:16.2$ & $0.44^{0.10}_{0.10}$ & $0.52^{0.36}_{0.28}$ & $0.37^{0.22}_{0.19}$ & $0.36^{0.19}_{0.17}$ \\ 
${\rm~HDFS1-334}$ & $32:52.91$ & $33:16.9$ & $1.28^{0.16}_{0.16}$ & $5.56^{2.53}_{2.58}$ & $4.34^{1.95}_{1.71}$ & $4.45^{1.99}_{1.44}$ \\ 
${\rm~HDFS1-340}$ & $32:55.90$ & $33:17.6$ & $0.52^{0.11}_{0.11}$ & $0.32^{0.20}_{0.15}$ & $0.21^{0.11}_{0.10}$ & $0.18^{0.09}_{0.08}$ \\ 
${\rm~HDFS1-342}$ & $33:0.18$ & $33:18.7$ & $0.74^{0.12}_{0.12}$ & $0.62^{0.26}_{0.25}$ & $0.37^{0.15}_{0.14}$ & $0.31^{0.12}_{0.11}$ \\ 
${\rm~HDFS1-346}$ & $32:54.31$ & $33:20.2$ & $0.46^{0.10}_{0.11}$ & $0.19^{0.13}_{0.10}$ & $0.12^{0.07}_{0.06}$ & $0.10^{0.05}_{0.05}$ \\ 
${\rm~HDFS1-347}$ & $32:53.12$ & $33:20.2$ & $3.28^{0.31}_{0.33}$ & $11.08^{1.64}_{1.60}$ & $5.83^{0.82}_{0.48}$ & $4.92^{1.15}_{0.36}$ \\ 
${\rm~HDFS1-345}$ & $33:2.81$ & $33:22.0$ & $0.56^{0.11}_{0.11}$ & $2.29^{1.87}_{1.41}$ & $2.09^{1.32}_{1.09}$ & $2.43^{1.22}_{1.09}$ \\ 
${\rm~HDFS1-350}$ & $33:5.00$ & $33:22.0$ & $3.04^{0.33}_{1.27}$ & $9.39^{3.24}_{7.74}$ & $7.60^{2.59}_{6.21}$ & $8.04^{2.74}_{6.53}$ \\ 
${\rm~HDFS1-355}$ & $32:54.24$ & $33:22.3$ & $2.88^{0.34}_{1.08}$ & $3.61^{1.44}_{3.00}$ & $2.80^{1.05}_{2.25}$ & $2.88^{1.09}_{2.20}$ \\ 
${\rm~HDFS1-354}$ & $32:57.26$ & $33:23.0$ & $1.38^{0.24}_{0.19}$ & $0.55^{0.38}_{0.24}$ & $0.42^{0.27}_{0.17}$ & $0.44^{0.27}_{0.17}$ \\ 
${\rm~HDFS1-364}$ & $32:57.08$ & $33:23.0$ & $0.68^{0.12}_{0.12}$ & $0.94^{0.49}_{0.35}$ & $0.53^{0.22}_{0.22}$ & $0.43^{0.18}_{0.16}$ \\ 
${\rm~HDFS1-363}$ & $32:52.15$ & $33:23.8$ & $1.12^{0.15}_{0.15}$ & $3.19^{1.26}_{1.22}$ & $2.01^{0.84}_{0.77}$ & $1.75^{0.77}_{0.66}$ \\ 
${\rm~HDFS1-360}$ & $33:2.88$ & $33:25.2$ & $1.30^{0.16}_{0.16}$ & $5.26^{1.82}_{1.92}$ & $3.52^{1.23}_{1.26}$ & $3.25^{1.15}_{1.13}$ \\ 
${\rm~HDFS1-368}$ & $33:0.94$ & $33:25.6$ & $0.96^{0.14}_{0.14}$ & $0.99^{0.49}_{0.36}$ & $0.55^{0.29}_{0.20}$ & $0.44^{0.26}_{0.17}$ \\ 
${\rm~HDFS1-372}$ & $32:50.57$ & $33:25.9$ & $0.56^{0.11}_{0.11}$ & $0.49^{0.29}_{0.24}$ & $0.33^{0.17}_{0.16}$ & $0.32^{0.15}_{0.14}$ \\ 
${\rm~HDFS1-373}$ & $32:50.71$ & $33:25.9$ & $0.54^{0.12}_{0.11}$ & $0.30^{0.19}_{0.15}$ & $0.17^{0.12}_{0.08}$ & $0.16^{0.10}_{0.07}$ \\ 
${\rm~HDFS1-378}$ & $32:50.68$ & $33:28.4$ & $2.62^{0.25}_{0.38}$ & $19.36^{4.57}_{5.42}$ & $11.99^{2.83}_{3.78}$ & $10.14^{2.39}_{2.98}$ \\ 
${\rm~HDFS1-379}$ & $32:53.05$ & $33:28.4$ & $1.06^{0.14}_{0.15}$ & $3.15^{2.09}_{1.41}$ & $2.82^{1.46}_{1.28}$ & $3.31^{1.33}_{1.38}$ \\ 
${\rm~HDFS1-377}$ & $32:55.00$ & $33:28.8$ & $1.12^{0.15}_{0.15}$ & $1.37^{0.95}_{0.65}$ & $1.46^{0.70}_{0.63}$ & $1.89^{0.66}_{0.75}$ \\ 
${\rm~HDFS1-380}$ & $32:57.12$ & $33:28.8$ & $0.68^{0.12}_{0.12}$ & $0.11^{0.09}_{0.06}$ & $0.12^{0.08}_{0.06}$ & $0.16^{0.09}_{0.08}$ \\ 
${\rm~HDFS1-381}$ & $32:59.50$ & $33:28.8$ & $1.00^{0.14}_{0.15}$ & $0.68^{0.30}_{0.24}$ & $0.40^{0.20}_{0.15}$ & $0.33^{0.18}_{0.13}$ \\ 
${\rm~HDFS1-382}$ & $32:58.31$ & $33:29.2$ & $2.62^{0.26}_{0.26}$ & $5.48^{1.41}_{1.19}$ & $3.44^{0.90}_{0.74}$ & $2.95^{0.79}_{0.62}$ \\ 
${\rm~HDFS1-386}$ & $33:3.24$ & $33:29.5$ & $2.64^{0.25}_{0.26}$ & $22.52^{5.21}_{4.13}$ & $13.77^{3.17}_{2.96}$ & $11.50^{2.63}_{2.22}$ \\ 
${\rm~HDFS1-383}$ & $32:58.24$ & $33:31.3$ & $0.42^{0.10}_{0.10}$ & $0.48^{0.33}_{0.25}$ & $0.31^{0.18}_{0.16}$ & $0.28^{0.15}_{0.14}$ \\ 
${\rm~HDFS1-424}\tablenotemark{b}$ & $32:56.83$ & $33:31.7$ & $4.82^{0.41}_{0.41}$ & $28.02^{10.54}_{6.91}$ & $24.97^{6.74}_{11.23}$ & $31.13^{7.16}_{18.65}$ \\ 
${\rm~HDFS1-393}$ & $33:1.80$ & $33:31.7$ & $1.62^{0.20}_{0.19}$ & $4.21^{1.52}_{1.35}$ & $2.96^{1.07}_{1.02}$ & $2.85^{1.04}_{1.00}$ \\ 
${\rm~HDFS1-394}$ & $33:4.28$ & $33:31.7$ & $0.10^{0.35}_{0.08}$ & $0.00^{0.13}_{0.00}$ & $0.00^{0.09}_{0.00}$ & $0.00^{0.08}_{0.00}$ \\ 
${\rm~HDFS1-395}$ & $32:54.71$ & $33:33.1$ & $0.16^{0.08}_{0.08}$ & $0.07^{0.12}_{0.05}$ & $0.05^{0.08}_{0.04}$ & $0.05^{0.08}_{0.04}$ \\ 
${\rm~HDFS1-397}$ & $32:53.41$ & $33:33.1$ & $1.10^{0.22}_{0.15}$ & $1.20^{1.33}_{0.50}$ & $1.00^{0.88}_{0.42}$ & $1.11^{0.78}_{0.41}$ \\ 
${\rm~HDFS1-399}$ & $32:52.37$ & $33:33.1$ & $0.52^{0.11}_{0.11}$ & $0.37^{0.24}_{0.18}$ & $0.24^{0.12}_{0.12}$ & $0.22^{0.11}_{0.10}$ \\ 
${\rm~HDFS1-404}$ & $32:55.75$ & $33:33.5$ & $0.54^{0.11}_{0.11}$ & $0.24^{0.14}_{0.10}$ & $0.15^{0.07}_{0.07}$ & $0.12^{0.06}_{0.05}$ \\ 
${\rm~HDFS1-405}$ & $33:0.04$ & $33:33.8$ & $1.02^{0.14}_{0.14}$ & $0.30^{0.23}_{0.14}$ & $0.36^{0.19}_{0.16}$ & $0.49^{0.20}_{0.21}$ \\ 
${\rm~HDFS1-398}$ & $32:53.30$ & $33:34.9$ & $0.96^{0.17}_{0.16}\tablenotemark{a}$ & $0.17^{0.18}_{0.10}$ & $0.18^{0.14}_{0.09}$ & $0.23^{0.14}_{0.11}$ \\ 
${\rm~HDFS1-406}$ & $32:47.65$ & $33:36.0$ & $0.58^{0.11}_{0.11}$ & $4.83^{3.83}_{2.83}$ & $4.38^{2.70}_{2.23}$ & $5.08^{2.49}_{2.23}$ \\ 
${\rm~HDFS1-411}$ & $32:54.96$ & $33:36.7$ & $1.00^{0.15}_{0.14}$ & $0.30^{0.22}_{0.12}$ & $0.26^{0.15}_{0.11}$ & $0.31^{0.14}_{0.13}$ \\ 
${\rm~HDFS1-427}$ & $33:2.88$ & $33:37.1$ & $1.18^{0.23}_{0.18}$ & $0.59^{0.54}_{0.24}$ & $0.44^{0.34}_{0.20}$ & $0.46^{0.30}_{0.19}$ \\ 
${\rm~HDFS1-414}$ & $32:51.50$ & $33:37.4$ & $0.62^{0.11}_{0.11}$ & $0.86^{0.50}_{0.44}$ & $0.65^{0.33}_{0.30}$ & $0.66^{0.31}_{0.27}$ \\ 
${\rm~HDFS1-410}$ & $32:53.77$ & $33:37.4$ & $0.52^{0.11}_{0.11}$ & $1.03^{0.67}_{0.49}$ & $0.66^{0.33}_{0.33}$ & $0.59^{0.31}_{0.26}$ \\ 
${\rm~HDFS1-415}$ & $32:59.46$ & $33:39.6$ & $0.46^{0.10}_{0.10}$ & $0.49^{0.29}_{0.25}$ & $0.32^{0.17}_{0.16}$ & $0.29^{0.15}_{0.13}$ \\ 
${\rm~HDFS1-421}$ & $33:3.64$ & $33:41.4$ & $0.44^{0.10}_{0.10}$ & $1.76^{1.43}_{1.00}$ & $1.43^{0.95}_{0.77}$ & $1.55^{0.82}_{0.77}$ \\ 
${\rm~HDFS1-426}$ & $32:54.02$ & $33:41.4$ & $1.00^{0.14}_{0.15}$ & $0.59^{0.39}_{0.27}$ & $0.48^{0.27}_{0.21}$ & $0.56^{0.23}_{0.24}$ \\ 
${\rm~HDFS1-434}$ & $32:49.45$ & $33:43.9$ & $0.58^{0.12}_{0.11}$ & $0.13^{0.10}_{0.06}$ & $0.09^{0.06}_{0.04}$ & $0.09^{0.05}_{0.04}$ \\ 
${\rm~HDFS1-435}$ & $32:47.47$ & $33:44.3$ & $0.56^{0.11}_{0.11}$ & $0.10^{0.07}_{0.05}$ & $0.08^{0.05}_{0.04}$ & $0.08^{0.04}_{0.04}$ \\ 
${\rm~HDFS1-437}$ & $32:49.99$ & $33:45.0$ & $1.06^{0.14}_{0.15}$ & $1.35^{0.56}_{0.42}$ & $0.74^{0.26}_{0.30}$ & $0.58^{0.25}_{0.23}$ \\ 
${\rm~HDFS1-439}$ & $33:2.52$ & $33:46.4$ & $0.68^{0.12}_{0.12}$ & $1.38^{0.66}_{0.59}$ & $0.83^{0.35}_{0.34}$ & $0.68^{0.29}_{0.26}$ \\ 
${\rm~HDFS1-440}$ & $32:58.63$ & $33:46.4$ & $1.34^{0.16}_{0.17}$ & $0.82^{0.50}_{0.33}$ & $1.03^{0.47}_{0.39}$ & $1.41^{0.55}_{0.53}$ \\ 
${\rm~HDFS1-448}$ & $32:45.56$ & $33:47.2$ & $1.30^{0.16}_{0.16}$ & $2.35^{0.59}_{0.74}$ & $1.34^{0.35}_{0.40}$ & $1.08^{0.27}_{0.33}$ \\ 
${\rm~HDFS1-450}$ & $32:57.88$ & $33:49.0$ & $0.44^{0.10}_{0.10}$ & $0.11^{0.07}_{0.06}$ & $0.07^{0.04}_{0.04}$ & $0.07^{0.03}_{0.03}$ \\ 
${\rm~HDFS1-463}$ & $33:3.10$ & $33:53.3$ & $2.76^{0.28}_{0.50}$ & $4.26^{1.25}_{1.22}$ & $3.31^{1.07}_{1.24}$ & $3.53^{1.25}_{1.55}$ \\ 
${\rm~HDFS1-484}$ & $32:46.90$ & $33:54.7$ & $0.52^{0.11}_{0.11}$ & $0.55^{0.56}_{0.32}$ & $0.63^{0.45}_{0.33}$ & $0.85^{0.49}_{0.43}$ \\ 
${\rm~HDFS1-472}$ & $32:48.26$ & $33:55.1$ & $0.66^{0.12}_{0.12}$ & $0.36^{0.20}_{0.17}$ & $0.25^{0.12}_{0.11}$ & $0.24^{0.11}_{0.10}$ \\ 
${\rm~HDFS1-476}$ & $33:0.90$ & $33:56.9$ & $1.08^{0.15}_{0.15}$ & $1.03^{0.41}_{0.37}$ & $0.61^{0.26}_{0.23}$ & $0.50^{0.23}_{0.20}$ \\ 
${\rm~HDFS1-480}$ & $32:53.02$ & $33:56.9$ & $2.76^{0.53}_{0.66}\tablenotemark{a}$ & $1.43^{1.54}_{0.85}$ & $1.79^{1.73}_{1.06}$ & $2.47^{2.25}_{1.46}$ \\ 
${\rm~HDFS1-479}$ & $32:59.24$ & $33:57.2$ & $1.34^{1.77}_{0.17}\tablenotemark{a}$ & $1.16^{11.35}_{0.48}$ & $0.84^{8.05}_{0.32}$ & $0.81^{7.82}_{0.27}$ \\ 
${\rm~HDFS1-483}$ & $33:2.74$ & $33:58.0$ & $2.24^{0.31}_{0.23}$ & $6.19^{1.90}_{1.36}$ & $3.84^{1.27}_{0.79}$ & $3.37^{1.05}_{0.67}$ \\ 
${\rm~HDFS1-487}$ & $32:51.54$ & $33:58.3$ & $1.28^{0.16}_{0.24}\tablenotemark{a}$ & $0.28^{0.15}_{0.17}$ & $0.27^{0.12}_{0.14}$ & $0.32^{0.13}_{0.16}$ \\ 
${\rm~HDFS1-488}$ & $32:52.15$ & $33:59.4$ & $0.48^{0.14}_{0.10}$ & $0.44^{0.53}_{0.23}$ & $0.35^{0.34}_{0.18}$ & $0.37^{0.29}_{0.17}$ \\ 
${\rm~HDFS1-492}$ & $32:51.32$ & $34:01.6$ & $0.24^{0.90}_{0.12}\tablenotemark{a}$ & $0.02^{0.81}_{0.02}$ & $0.01^{0.46}_{0.01}$ & $0.01^{0.36}_{0.01}$ \\ 
${\rm~HDFS1-489}$ & $32:52.26$ & $34:02.6$ & $0.52^{0.11}_{0.11}$ & $0.48^{0.37}_{0.26}$ & $0.37^{0.24}_{0.20}$ & $0.40^{0.20}_{0.19}$ \\ 
${\rm~HDFS1-478}$ & $32:50.96$ & $34:04.8$ & $1.34^{0.16}_{0.17}$ & $5.12^{1.83}_{2.03}$ & $3.37^{1.36}_{1.19}$ & $3.17^{1.27}_{1.03}$ \\ 
${\rm~HDFS1-505}$ & $32:59.86$ & $34:05.5$ & $1.30^{0.16}_{0.17}$ & $1.92^{0.76}_{0.76}$ & $1.33^{0.54}_{0.52}$ & $1.27^{0.52}_{0.48}$ \\ 
${\rm~HDFS1-511}$ & $32:49.85$ & $34:06.2$ & $1.12^{0.15}_{0.15}$ & $2.43^{1.11}_{1.01}$ & $1.64^{0.79}_{0.64}$ & $1.52^{0.78}_{0.53}$ \\ 
${\rm~HDFS1-516}$ & $32:55.28$ & $34:07.7$ & $0.46^{0.10}_{0.10}$ & $0.33^{0.19}_{0.16}$ & $0.20^{0.10}_{0.09}$ & $0.16^{0.08}_{0.07}$ \\ 
${\rm~HDFS1-542}$ & $32:51.11$ & $34:08.0$ & $3.86^{0.34}_{3.46}$ & $31.10^{6.03}_{31.03}$ & $15.42^{1.61}_{15.36}$ & $12.60^{1.02}_{12.55}$ \\ 
${\rm~HDFS1-521}$ & $32:47.58$ & $34:08.8$ & $0.50^{0.11}_{0.11}$ & $0.89^{0.67}_{0.48}$ & $0.67^{0.43}_{0.36}$ & $0.69^{0.36}_{0.33}$ \\ 
${\rm~HDFS1-522}$ & $33:4.50$ & $34:08.8$ & $0.56^{0.11}_{0.11}$ & $0.15^{0.09}_{0.07}$ & $0.10^{0.05}_{0.05}$ & $0.09^{0.04}_{0.04}$ \\ 
${\rm~HDFS1-530}$ & $32:55.25$ & $34:10.2$ & $1.02^{0.14}_{0.14}$ & $1.36^{0.60}_{0.52}$ & $0.90^{0.37}_{0.37}$ & $0.83^{0.36}_{0.32}$ \\ 
${\rm~HDFS1-536}$ & $33:1.58$ & $34:10.6$ & $0.78^{0.13}_{0.13}$ & $0.58^{0.28}_{0.25}$ & $0.37^{0.18}_{0.15}$ & $0.33^{0.16}_{0.13}$ \\ 
${\rm~HDFS1-527}$ & $33:1.80$ & $34:13.4$ & $1.12^{0.15}_{0.15}$ & $3.83^{2.51}_{1.67}$ & $3.59^{1.74}_{1.55}$ & $4.33^{1.56}_{1.71}$ \\ 
${\rm~HDFS1-538}$ & $32:56.11$ & $34:14.2$ & $0.52^{0.11}_{0.11}$ & $0.66^{0.43}_{0.30}$ & $0.42^{0.22}_{0.21}$ & $0.37^{0.19}_{0.16}$ \\ 
${\rm~HDFS1-548}$ & $33:0.54$ & $34:17.4$ & $0.66^{0.12}_{0.12}$ & $0.19^{0.17}_{0.10}$ & $0.23^{0.15}_{0.11}$ & $0.30^{0.17}_{0.14}$ \\ 
${\rm~HDFS1-555}$ & $32:59.60$ & $34:20.3$ & $1.12^{0.15}_{0.15}$ & $0.98^{0.63}_{0.51}$ & $1.02^{0.50}_{0.43}$ & $1.28^{0.52}_{0.46}$ \\ 
\label{tbl:table3.in}
\enddata
\tablenotetext{a} {$\geq1\%$ of Monte-Carlo realizations have z more than unity away from $z_{phot}$}
\tablenotetext{b} {$z_{phot}$ may be discrepant}
\end{deluxetable}

\clearpage

\end{document}